\documentclass[twocolumn,showpacs,preprintnumbers,superscriptaddress,amsmath,amssymb,nofootinbib, floatfix]{revtex4}
\usepackage{epsfig,graphics,color,graphicx,amsmath}
\usepackage[section]{placeins}

\usepackage[usenames,dvipsnames]{xcolor}
\usepackage[normalem]{ulem}

\begin{document}
\title{Solving Bethe-Salpeter scattering state equation in Minkowski space
}

\author{J.~Carbonell}
\affiliation {Institut de Physique Nucl\'eaire,
Universit\'e Paris-Sud, IN2P3-CNRS, 91406 Orsay Cedex, France}
\author{V.A. Karmanov}
\affiliation {Lebedev Physical Institute, Leninsky Prospekt 53, 119991
Moscow, Russia}

\bibliographystyle{unsrt}

\begin{abstract}
We present a method to directly solving  the Bethe-Salpeter equation in Minkowski space, both for bound and scattering states.
It is based on a proper treatment of the singularities which appear in the kernel, propagators and Bethe-Salpeter amplitude itself.  
The off-mass shell scattering amplitude for spinless particles interacting by a one boson exchange   is computed for the first time. 
\end{abstract}
\pacs{03.65.Pm, 03.65.Ge, 11.10.St}

\maketitle
%

\section{Introduction} \label{intro}

The interest  in  solving the  Bethe-Salpeter (BS) equation \cite{bs}  in its natural Minkowski-space formulation has  increased in the recent years
\cite{KusPRD95,KusPRD97,bs1,bs2,ck-trento,Sauli_JPG08,ckm_ejpa,fsv-2012,CK_PLB_2013,lc2012,HL_PRD85_2012,Fukuoka,LC2013-2,transit,FrePRD14,BGTA_FBS54_2013,BGTA_PRD89_2014}.
There are several  reasons for dealing with Minkowski solutions. 
One of them is the fact that the Wick rotation is not directly applicable
for computing electromagnetic form factors due to the singularities in the complex momentum plane \cite{ckm_ejpa,ck-trento}.
The Euclidean solutions are still used  in the context of BS-Schwinger-Dyson equations for computing bound-state form factors but 
this requires a huge numerical effort
\cite{Oettel,Bhagwat:2002tx,Maris:2005tt,Krassnigg:2009gd,Eichmann,Strauss:2012dg,Windisch:2012sz,ChaPRL13}.
They can also be used to obtain on-shell observables  like binding energies or phase shifts \cite{tjon,Schwartz_Morris_Haymaker},
whereas  the computation of the off-shell BS scattering amplitude  --
mandatory for computing  e.g. the transition e.m. form factor or  for solving the three- and many-body  BS equations --
is possible only using a full Minkowski solution.  

A method for computing these solutions  based on the  Nakanishi  representation  \cite{nak63} of the BS amplitude was first developed in  \cite{KusPRD95,KusPRD97}.
A similar approach combined with the light-front projection  was proposed in \cite{bs1,bs2}. 
It led to a different integral equation  which involved only smooth functions 
and was numerically easy to treat. The bound state Minkowski amplitude and later on \cite{ckm_ejpa,ck-trento} the corresponding  form factors
were in this way computed for the first time.
A modified method  to the one developed in \cite{bs1,bs2} aimed to compute the scattering states was proposed in \cite{fsv-2012}. 
It has already been  successfully  tested for the bound states \cite{FrePRD14}.

Although our approach  \cite{bs1,bs2} could also be naturally extended to the scattering states,
we have developed a new method \cite{CK_PLB_2013,lc2012,Fukuoka,LC2013-2}  which  allows to solve the Minkowski BS equation 
 in a simplest and more straightforward way.
 It consists in a direct solution of the equation which takes properly into account the many singularities  and without making use of the Nakanishi integral representation.
The aim of this paper is to present  this method in detail with applications to the problem of
two scalar particles interacting  by a one-boson exchange kernel.

Some of the results   
have been presented in the short  publications \cite{CK_PLB_2013,lc2012,Fukuoka} and reviewed in \cite{LC2013-2} without a detailed explanation of the method.
Until now, the off-shell BS amplitude has been computed only for a separable kernel \cite{burov}.

In Sec. \ref{transform} we transform the Bethe-Salpeter equation to the form which does not contain the pole singularities. 
In Sec. \ref{kernel} we analyze the kernel singularities.
Section \ref{phase} is devoted to the extraction of the phase shifts from the computed Minkowski amplitude.
We derive in Sec. \ref{euclid_scat}  the system of equations which couples the Euclidean amplitude with the Minkowski one for a particular value of its arguments. The comparison between the direct solution in Minkowski space and the one found using this system of equations constitutes a strong test for 
our approach. The numerical results are presented in Sec. \ref{num}. 
They concern the half-off-shell BS amplitude,  the scattering length and the elastic and inelastic phase shifts.
 Sec. \ref{concl} contains some concluding remarks. Technical details are given in appendices \ref{Numerics}, \ref{appen1} and \ref{first}.

\section{Transforming the BS equation}\label{transform}

Let us consider the scattering of two equal mass ($m$) particles  with initial ($k_{is}$)  and final ($k_i$)  four-momenta  respectively 
\[  k_{1s}+ k_{2s}  \rightarrow     k_1+ k_2     \]  
The corresponding BS  amplitude  $F$ is parametrized in terms of the total 
\begin{equation}\label{k_i}
 p=k_1+ k_2 = k_{1s}+ k_{2s}  
 \end{equation}
and relative momenta
 \begin{eqnarray} 
 2k    &=&k_1- k_2   \cr
 2k_s&=& k_{1s}- k_{2s}
 \end{eqnarray}
 The subscript $s$ means "scattering" (on-mass-shell) momenta.
For a scattering process, $F$  obeys the inhomogeneous integral equation graphically represented in Fig. \ref{bs_eq}.
\begin{widetext}
\begin{center}
\begin{figure}[hbtp]
\centering
\includegraphics[width=12.cm]{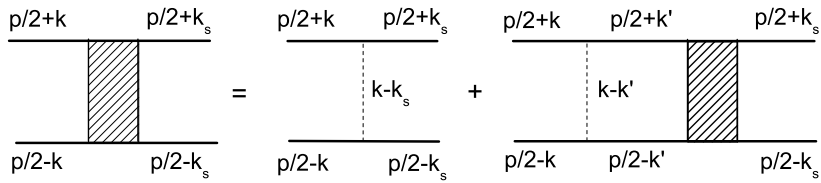}
\caption{Bethe-Salpeter equation for a scattering state.}\label{bs_eq}
\end{figure}
\end{center}
\end{widetext}
Its analytic expression in Minkowski space reads:
\begin{widetext}
\begin{equation}\label{BSE}
F(k,k_s;p)=K(k,k_s;p)- i\int\frac{d^4k'}{(2\pi)^4} \frac{K(k,k';p) F(k',k_s;p)}
{\left[\left(\frac{p}{2}+k'\right)^2-m^2+i\epsilon\right]  \left[\left(\frac{p}{2}-k'\right)^2-m^2+i\epsilon\right]} 
\end{equation}
\end{widetext}
We will  consider all along this article the case of spinless particles  interacting by the one-boson exchange kernel  $K$:
\begin{equation}\label{obe}
K(k,k';p)=-\frac{16\pi m^2\alpha}{(k-k')^2-\mu^2+i\epsilon},
\end{equation}
where \mbox{$\alpha=g^2/(16\pi m^2)$}  is the dimensionless coupling constant.
In the non relativistic limit, this kernel leads to the Yukawa potential \mbox{$V(r)=-\alpha\exp(-\mu r)/r$}.
We denote by  $M^2=p^2$,  the squared total invariant mass of the system.

\bigskip
The amplitude $F$ depends on the three four-momenta $k,k_s,p$. 
In the  center of mass frame, defined by $\vec{p}=0$, one has 
$k_s=(0,\vec{k}_s)$  and \mbox{$p_0=M=2\varepsilon_{k_s}=2\sqrt{m^2+k_s^2}$.}
For a given incident  momentum $\vec{k}_s$,
$F$  depends  on the three scalar variables  $k_0$, $|\vec{k}|$  and $z=\cos(\vec{k},\vec{k}_s)$.
It will be hereafter denoted  by $F(k_0,k,z)$, setting abusively $k=|\vec{k}|$, $k_s=|\vec{k_s}|$.
The modulus of the incident momentum $k_s$ plays the role of a parameter (like the bound state mass $M$ in the bound state equation)
and therefore it will not be included in the arguments of the amplitude.
However, in contrast to the bound state case, one has $M>2m$ and $F$ depends also on the extra variable $z=\cos \theta$, where $\theta$ is the scattering angle.

Notice that the solution thus obtained is the  half-off-mass shell amplitude. 
It  is a particular case of the so called full off-shell amplitude $F(k_0,k,z;k_{0s},k_s;M)$.  
The latter, in addition to the variables $k_0,k,z$, depends also on the off-shell independent variables $k_{0s},k_s$, 
now with $k_{0s}\neq 0$ and $k_{0s}\neq \varepsilon_{k_s}$ and the total mass $M$ which is an independent parameter  neither equal to $2\varepsilon_{k_s}$ nor related to $k_{0s}$. 
By "off-shell amplitude" we will hereafter mean the half-off-shell amplitude. 
The method we have developed  can also be applied to the full off-shell amplitude, 
though the dependence of the latter on two extra variables $k_{0s},k_s$ requires much more extensive numerical calculations and will not be considered here.

\bigskip
The difficulty in computing  the off-shell amplitude $F(k_0, k, z)$ in the entire domain of its arguments is due to the singular character of 
the inhomogeneous term $K$ and  	as well as of each of the 
factors in the integrand of  equation  (\ref{BSE}).
In particular the  singular character of the amplitude $F$ itself makes it hardly representable in terms of smooth functions.
These singularities are integrable in the mathematical sense, due to $i\epsilon$ in the denominators of propagators,
but  their  integration is a quite delicate task and requires the use of appropriate analytical as well as  numerical methods.

To avoid these problems, equation (\ref{BSE}) was first solved on-shell  \cite{tjon}  by rotating the integration contour $k_0\to ik_4$ and taking into account
the contributions of the crossed singularities. These singularities are absent in the bound state case but exist for the scattering states.
A similar method will be developed in Sec. \ref{euclid_scat} as a test of our approach. 

\bigskip
The off-shell amplitude $F(k_0,k,z)$ can be obtained by directly solving the corresponding three-dimensional equation
derived from (\ref{BSE}) after integrating over the azimuthal variable.
However we prefer to present in what follows its partial wave solution.
This procedure, apart from the much smaller numerical cost, has the advantage of smoothing the kernel singularities in particular in the inhomogeneous term. 
The partial wave amplitude $F_L(k_0,k)$ is defined as  \cite{IZ}:
\begin{equation}\label{fpw1}
F(k_0,k,z)=16\pi\sum_{L=0}^{\infty}(2L+1) F_L(k_0,k)P_L(z)
\end{equation}
where $P_L(z)$ is the Legendre polynomial and
\begin{equation}\label{fpw2}
 F_L(k_0,k)=\frac{1}{32\pi}\int_{-1}^1dz \; P_L(z)F(k_0,k,z)
 \end{equation}

By  inserting  (\ref{fpw1}) into  (\ref{BSE}) we  obtain a set of uncoupled two dimensional equations for the partial amplitudes $F_L$. 
\begin{widetext}
\begin{equation}\label{BSE1}
F_L(k_0,k)=F_L^B(k_0,k)- i\int_0^{\infty} {k'}^2dk'\int_{-\infty}^{\infty}dk'_0 
\frac{W_L(k_0,k,k'_0,k') F_L(k'_0,k')}{(k'_0-a_- +i\epsilon)(k'_0+a_- -i\epsilon) (k'_0-a_+ +i\epsilon)(k'_0+a_+ -i\epsilon)}
\end{equation}
\end{widetext}
with
\[ W_L(k_0,k,k'_0,k')=  \frac{1}{(2\pi)^3} \int_{-1}^1 dz\; P_L(z)  K(k,k';p)  \]
and the inhomogeneous (Born)  term $F^B_L$ is given in terms of $W_L$ by:
\begin{equation}\label{FB}
F^B_L(k_0,k)={\pi^2\over 4}W_L(k_0,k,k'_0=0,k'=k_s)
\end{equation}

The denominator of (\ref{BSE})   has been factorized   by writting (in the c.m. frame).  
\begin{eqnarray*}
\left(\frac{p}{2}+k'\right)^2-m^2+i\epsilon  &=& \left( \varepsilon_{k_s}  + k'_0\right)^2 -  (\varepsilon_{k'}-i\epsilon)^2   \cr
\left(\frac{p}{2}-k'\right)^2-m^2+i\epsilon   &=& \left( \varepsilon_{k_s}   - k'_0\right)^2 -  (\varepsilon_{k'}-i\epsilon)^2 
\end{eqnarray*}
and making the replacement \mbox{$ -\varepsilon_{k'}^2+i\epsilon\to -(\varepsilon_{k'}-i\epsilon)^2 $,} valid since it does not change the sign of imaginary contribution.
This leads to the expression displayed in (\ref{BSE1})  where the four propagator poles  are made explicit. They are symmetric with respect the origin in the complex $k'_0$-plane  and are given by 
\begin{eqnarray}
{k'}^{(1)}_0&=&\phantom{-}\varepsilon_{k_s}+\varepsilon_{k'} - i\epsilon = +a_+  -   i\epsilon            \cr
{k'}^{(2)}_0&=&\phantom{-}\varepsilon_{k_s}-\varepsilon_{k'}+ i\epsilon  = -a_-    +  i\epsilon      \cr
{k'}^{(3)}_0&=&-\varepsilon_{k_s}+\varepsilon_{k'}- i\epsilon  = +a_- - i\epsilon      \cr
{k'}^{(4)}_0&=&-\varepsilon_{k_s}-\varepsilon_{k'}+ i\epsilon  = - a_+ + i\epsilon   \label{Poles}
\end{eqnarray}
with
\begin{equation}\label{a+-}
a_{\pm} =\varepsilon_{k'}   \pm \varepsilon_{k_s} 
\end{equation}
Notice that $a_+>0$ while for the scattering process $a_-$ vs. $k'$ changes sign and $a_-(k'=k_s)=0$.

We will be hereafter restricted the Bethe-Salpeter solutions for  \mbox{S-wave}.
The corresponding kernel $W_0$ is given by:

\begin{widetext}
\begin{eqnarray}\label{W0}
W_0(k_0,k,k'_0,k')  & \equiv &   {1\over kk'} w_0(\eta)  = -\frac{\alpha m^2}{\pi kk'} \left\{  \frac{1}{\pi}  \log \left| \frac{(\eta+1)}{(\eta-1)} \right| - i I(\eta)   \right\}
\end{eqnarray}
with
\begin{equation}\label{eta}
I(\eta)=\left\{
\begin{array}{lcrcl}
1  & {\rm if} & \mid\eta\mid &\leq& 1 \cr
0  & {\rm if} & \mid\eta\mid &> & 1
\end{array}\right.
\qquad
\eta =   {(k_0  - k'_0)^2 - k^2 - {k'}^2 -\mu^2 \over2kk'} 
\end{equation}
\end{widetext}

\begin{figure}[hbtp]
\begin{center}
\includegraphics[width=7.5cm]{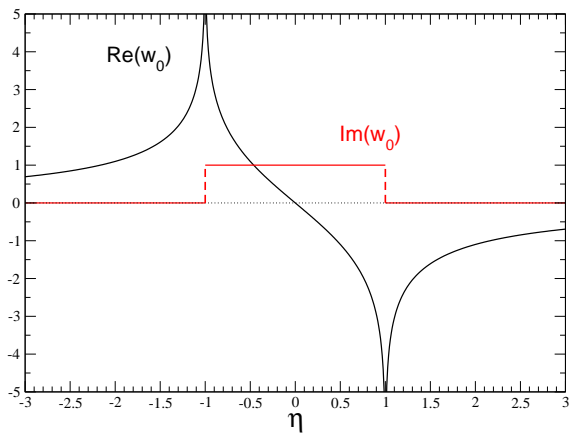}
\end{center}
\caption{S-wave reduced kernel $w_0$, Eq. (\ref{W0}), as a function of variable $\eta$ defined in (\ref{eta})}\label{Fig_W0}
\end{figure}

The reduced kernel $w_0(\eta)$ has singularities  both in its real and imaginary parts.
Its real part is an odd function of $\eta$ with logarithmic singularities at \mbox{$\eta=\pm1$,}
its imaginary part is an even function of $\eta$ with discontinuities  at the same points.
It is represented in Fig. \ref{Fig_W0} as a function of variable $\eta$.
\bigskip\bigskip

The solution of (\ref{BSE1}) faces three kind of problems, all related to the unavoidable singularities when working  in Minkowski metric,
and  must be properly treated before a numerical solution can be tried.

First are the four propagator  poles in the right hand side of (\ref{BSE1})  which are  explicitly given by (\ref{Poles}).

Second are the logarithmic singularities of the kernel 
$W_0$  which make difficult its numerical integration both in $k'_0$ and $k'$ variables. 

Third, are the singularities of the inhomogeneous term $F_0^B$. They are related to the previous ones, i.e. $W_0$, but generate a different type of problems:
they  imply the singular character of  the amplitude  $F_0$ we are interested in,  and thus a  difficulty in  being represented  in terms of smooth functions
when solving numerically equation (\ref{BSE1}).

In the following subsections  we will examine separately each of these points and detail our approach to circumvent the related difficulties.

\subsection{Removing the pole singularities}

Let us first  represent the pole contributions in the integrand of (\ref{BSE1}) in the usual form:
\begin{eqnarray*}
\frac{1}{k'_0 -a_{\pm}  + i\epsilon} &=& PV\left(\frac{1}{k'_0 -a_{\pm}}\right)   -  i\pi\delta(k'_0-a_{\pm})  , \\
\frac{1}{k'_0+a_{\pm}  -  i\epsilon} &=& PV\left(\frac{1}{k'_0+a_{\pm}}\right)  + i\pi\delta(k'_0+a_{\pm})
 \end{eqnarray*}
where PV denotes the principal value. 
The integrand  in the r.h.-side of Eq. (\ref{BSE1}) takes the form
\begin{widetext}
\begin{small}
\begin{eqnarray}
\frac{f(k_0,k,k'_0,k')}{(k'_0-a_+ + i\epsilon)(k'_0+ a_+ - i\epsilon)(k'_0+ a_- - i\epsilon)(k'_0-a_- + i\epsilon)}
&      =   &  \left[             -i\pi \delta(k'_0 -a_+)     + PV {1\over k'_0-a_+}   \right]  \cr
&\times &  \left[    i\pi \delta(k'_0+a_+)   + PV {1\over k'_0+a_+}   \right] \cr
&\times &  \left[     - i\pi \delta(k'_0 -a_-)     + PV {1\over k'_0-a_-}     \right] \cr
&\times &  \left[    i\pi \delta(k'_0+a_-)    + PV {1\over k'_0+a_-}    \right]        f(k_0,k,k'_0,k')  \label{Integrand}
\end{eqnarray}
\end{small}
\end{widetext}
with the notation
\begin{equation}\label{f}
f(k_0,k,k'_0,k') \equiv   W_0(k_0,k,k'_0,k') F_0(k'_0,k')
\end{equation}
By expanding this expression  we obtain the integral term on the r.h.-side of  (\ref{BSE1})
as a sum of terms containing respectively products of 4,3,2,1 and 0 delta-functions to be integrated over $k'_0$ and $k'$ variables:
\[I = I_4+I_3+I_2+I_1+I_0 \]

The terms $I_4$ and $I_3$ containing a products of four and three  delta-functions  are always zero since
the arguments of $\delta$'s cannot vanish simultaneously.

Among the  terms $I_2$ containig the product of  two delta-functions, and for the same reasons than in the previous case,
only the one containing  \mbox{$\delta(k'_0-a_-)\delta(k'_0+a_-)$}  gives a 
non zero contribution to the integral  when $k'_0=\pm a_-=0$,
that is when $\varepsilon_{k'} = \varepsilon_{k_s}$. 
This contribution reads:

\begin{widetext}
\begin{eqnarray}
 I_2 &=&  -i\pi^2 \int_0^{\infty} {k'}^2 \;  dk'   \int_{-\infty}^{+\infty} dk'_0   \delta(k'_0-a_-) \delta(k'_0+a_-) {f(k_0,k,k'_0,k' ) \over {k'}^2_0-a_+^2}  ÃÂ\cr
       &=& +i \pi^2 \int_0^{\infty} ÃÂ dk'     \delta[2 (  \varepsilon_{k'} - \varepsilon_{k_s}  ) ]  {1\over 4 \epsilon_{k'}  \epsilon_{k_s} }  f(k_0,k,0,k' ) 
       =   \frac{ i\pi^2k_s}{8\varepsilon_{k_s}} W_0(k_0,k,0,k_s) F_0(0,k_s)  \label{2delta}
 \end{eqnarray}
\end{widetext}
where we have used 
\begin{eqnarray}
  {1\over a_-^2-a_+^2}  &=&  - {1\over 4 \epsilon_{k'}  \epsilon_{k_s} }  \label{R1}\\
 \delta[ \varepsilon_{k'}-\varepsilon_{k_s}    ] &=&  {  \epsilon_{k_s} \over  k_s  } \delta (k'-k_s)  \label{R2}
 \end{eqnarray}

\bigskip
The sum of the four terms  from (\ref{Integrand}) containing one delta-function  reads:
\begin{widetext}
\begin{eqnarray}\label{I1}
I_1&=-i      & \int_0^{\infty} {k'}^2dk' \int_{-\infty}^{\infty}dk'_0\, f(k_0,k,k'_0,k')     \\
     &\times & \left[-i\pi\delta(k'_0-a_+)\frac{1}{ {k'_0}^2-a_-^2}PV\frac{1}{k'_0+a_+}+i\pi\delta(k'_0+a_+)\frac{1}{{k'_0}^2-a_-^2}PV\frac{1}{k'_0-a_+}\right.         \nonumber\\
    &             & \phantom{[}\left. -i\pi\delta(k'_0-a_-)\frac{1}{{k'_0}^2-a_+^2}PV\frac{1}{k'_0+a_-}+i\pi\delta(k'_0+a_-)\frac{1}{{k'_0}^2-a_+^2}PV\frac{1}{k'_0-a_-}\right]
\nonumber
\end{eqnarray}
After integrating over $dk'_0$, the two integrals over $dk'$ remain:
\begin{eqnarray}
I_1&=&\frac{\pi}{4\varepsilon_{k_s}}PV \int_0^{\infty}  \frac{{k'}^2 dk'}{2\varepsilon_{k'}a_-} [f(k_0,k,k'_0=a_-,k')+f(k_0,k,k'_0=-a_-,k')]    \cr        
     &-& \frac{\pi}{4\varepsilon_{k_s}} \phantom{PV} \int_0^{\infty} \frac{{k'}^2 dk'} {2\varepsilon_{k'}a_+}  [f(k_0,k,k'_0=a_+,k')+f(k_0,k,k'_0=-a_+,k')]  \label{I1a}
\end{eqnarray}
\end{widetext}

Since the $W_0$ kernel is symmetric with respect to the change of sign of variables $k_0,k'_0$
\[ W_0(-k_0,k,-k'_0,k')=W_0(k_0,k,k'_0,k') \]
the solution $F_0(k_0,k)$ is also symmetric, that is:  $F_0(-k_0,k)=F_0(k_0,k)$. 
By inserting this relation in (\ref{I1a}), one can see that the symmetrized value of the kernel $W_0$ with respect to the variable $k'_0$ 
\begin{widetext}
\begin{equation}\label{WS}
W_0^S(k_0,k,k'_0,k')=W_0(k_0,k,k'_0,k')+W_0(k_0,k,-k'_0,k')
\end{equation}
\end{widetext}
appears naturaly in the formulation.
After substituting  (\ref{f}) and (\ref{WS}) in  (\ref{I1a}) one gets: 
\begin{widetext}
\begin{equation}\label{I1b}
I_1=\frac{\pi}{4\varepsilon_{k_s}} PV\int_0^{\infty}  \frac{{k'}^2 dk'}{2\varepsilon_{k'}a_-} W_0^S(k_0,k,a_-,k')F_0(a_-,k')       
     - \frac{\pi}{4\varepsilon_{k_s}} \int_0^{\infty} \frac{{k'}^2 dk'} {2\varepsilon_{k'}a_+}W_0^S(k_0,k,a_+,k')F_0(a_+,k') 
\end{equation}
\end{widetext}

The first integrand in (\ref{I1b}) is  singular  due to the  factor $a_-=\varepsilon_{k'}-\varepsilon_{k_s}$ in the denominator which vanishes at $k'=k_s$. It must be understood in the sense of  principal value.
The integral is well defined but, because of the singularity of the integrand at \mbox{$k'=k_s$,} explicitly manifested in the form:
\[   \frac{1}{a_-}=\frac{1}{\varepsilon_{k'}-\varepsilon_{k_s}}=\frac{\varepsilon_{k'}+\varepsilon_{k_s}}{{k'}^2-k_s^2}  \]
it requires an additional treatment to be transformed  in a non-singular form. The singularity is eliminated using the subtraction technique, that is 
\begin{equation}\label{subtr}
PV\int_0^{\infty} \frac{h(k')dk'}{{k'}^2-a^2}=\int_0^{\infty} dk' \left[\frac{h(k')-h(a)}{{k'}^2-a^2}\right]
\end{equation}
based on the identity:
\begin{equation}\label{subtr1}
PV\int_{0}^{\infty} dk'\; \frac{1}{{k'}^2-a^2}=0,\quad \mbox{if $a\neq 0$.}
\end{equation}
The condition $a\neq0$  prevent us from setting $k_s=0$ in our equation. 
 
The second integrand in (\ref{I1b}) is non-singular and it does not require any additional treatment.

Let us finally consider in  (\ref{Integrand}) the term which does not contain any delta-function. Its contribution to r.h.-side of the equation  (\ref{BSE1})  can be represented as:

\begin{widetext}
\begin{eqnarray}\label{I0}
I_0&=&-i\int_0^{\infty}{k'}^2dk'\,PV\int_{-\infty}^{\infty} \frac{f(k_0,k,k'_0,k')dk'_0}{(k'_0-a_-)(k'_0+a_-)(k'_0-a_+)(k'_0+a_+)}        \nonumber \cr
     &=&-i\int_0^{\infty}\frac{{k'}^2dk'}{4\varepsilon_{k_s}\varepsilon_{k'} }\,PV\int_{-\infty}^{\infty} dk'_0 f(k_0,k,k'_0,k')  \left[\frac{1}{({k'}^2_0-a_+^2)}-\frac{1}{({k'}^2_0-a_-^2)}\right]       = I_0^+ + I_0^-
\end{eqnarray}
The singularity  $\frac{1}{({k'}^2_0-a_+^2)}$ in  $I_0^+$ is regularized by using the same subtraction technique (\ref{subtr}) than previously.
\begin{eqnarray*}
 I_0^+ &=&-i\int_0^{\infty}\frac{{k'}^2dk'}{4\varepsilon_{k_s}\varepsilon_{k'} }\,PV\int_{-\infty}^{\infty} dk'_0 \;   \frac{ f(k_0,k,k'_0,k')  }{({k'}^2_0-a_+^2)}   \cr
            &=& -i\int_0^{\infty}\frac{{k'}^2dk'}{4\varepsilon_{k_s}\varepsilon_{k'} }\,PV\int_{-\infty}^{\infty} dk'_0 \;   \left[ \frac{ f(k_0,k,k'_0,k')  }{({k'}^2_0-a_+^2)}   -\frac{ f(k_0,k,a_+,k')  }{({k'}^2_0-a_+^2)}      \right] 
 \end{eqnarray*}
\end{widetext}

The singularity $\frac{1}{({k'}^2_0-a_-^2)}$  in  $I_0^-$  requires some care since \mbox{$a_-=\varepsilon_{k'}-\varepsilon_{k_s}$} vanishes when $k'=k_s$ while the relation
(\ref{subtr}) is valid only if $a\neq 0$.
Notice that  (\ref{subtr1}) diverges if $a=0$ and therefore cannot be applied by the simple  replacement $k'\to k'_0$. 
We use instead the relation
\begin{equation}\label{subtr1aa}
PV\int_{-\infty}^{\infty}\frac{dx}{x^2-a^2}=\frac{\pi^2}{2}\delta(a)
\end{equation}
which has been derived in appendix \ref{appen1}.  For $a=a_-=\varepsilon_{k'}-\varepsilon_{k_s}$ it takes the form: 
\begin{equation}\label{subtr2}
PV\int_{-\infty}^{\infty}\frac{dk'_0}{[{k'}^2_0-(\varepsilon_{k'}-\varepsilon_{k_s})^2]}=\frac{\pi^2}{2}\delta(\varepsilon_{k'}-\varepsilon_{k_s})
\end{equation}
The subtraction formula (\ref{subtr}) must now be replaced by
\begin{widetext}
\begin{small}
\begin{eqnarray}
I_0^-  &=& i\int_0^{\infty}\frac{{k'}^2dk'}{4\varepsilon_{k_s}\varepsilon_{k'} }PV \int_{-\infty}^{\infty}dk'_0 \frac{f(k_0,k,k'_0,k')}{{k'_0}^2-a_-^2}  \cr
          &=&
i\int_0^{\infty}\frac{{k'}^2dk'}{4\varepsilon_{k_s}\varepsilon_{k'} } \int_{-\infty}^{\infty}dk'_0 \left[\frac{f(k_0,k,k'_0,k')}{{k'_0}^2-a_-^2} -\frac{f(k_0,k,a_-,k')}{{k'_0}^2-a_-^2}\right]
+ \frac{i\pi^2}{2}   \int_0^{\infty}  \frac{{k'}^2dk'}{4\varepsilon_{k_s}\varepsilon_{k'} }   \delta(\varepsilon_{k'}-\varepsilon_{k_s})      f(k_0,k,a_-,k')\label{subtr1a}
\end{eqnarray}
\end{small}
\end{widetext}
The integrand of the first term in  (\ref{subtr1a}) -- inside the square brackets --  is again regular. 

The last term is transformed using relation (\ref{R2}) and performing the integral over $k'$ into :
\begin{widetext}
\begin{equation}\label{nonint1}
\frac{i\pi^2}{2}\int_0^{\infty} \frac{{k'}^2dk'}{4\varepsilon_{k_s}\varepsilon_{k'} }  \delta(\varepsilon_{k'}-\varepsilon_{k_s})f(k_0,k,a_-,k') 
= \frac{{i\pi^2k_s}^2}{8\varepsilon_{k_s} } f(k_0,k,a_-,k_s) =
\frac{i{k_s}^2}{8\varepsilon_{k_s}^2}W_0(k_0,k,0,k_s)F_0(0,k_s)
\end{equation}
\end{widetext}
It gives exactly the same contribution as the two-delta term (\ref{2delta}). The sum of these two contributions
results into multiplying the coefficient in (\ref{2delta}) by a  factor 2.

Since, as noticed above, the solution $F_0(k_0,k)$ is symmetric with respect to $k_0\to -k_0$,
the equation (\ref{BSE1})  can  be reduced to the interval $k_0 \in[ 0,\infty]$.
After reducing the integral term to the same interval in $k'_0$ and introducing the symmetric kernel $W_S$  given by (\ref{WS}), we finally obtain the S-wave equation that we aimed to solve and that does not contain the pole singularities:
\begin{widetext}
\begin{eqnarray}
F_0(k_0,k)  &=& F^B_{0}(k_0,k)    +  \frac{i\pi^2 k_s}{8\varepsilon_{k_s}} W_0^S(k_0,k,0,k_s) F_0(0,k_s)
 \cr
                 &+& \frac{\pi}{2M} \int_0^{\infty}  \frac{dk'}{ \varepsilon_{k'} ( 2\varepsilon_{k'}-M) }    \left[{k'}^2 W_0^S(k_0,k, a_-,k') F_0(| a_- |,k')
     -\frac{2 {k_s}^2\varepsilon_{k'}}{\varepsilon_{k'} +\varepsilon_{k_s}}W_0^S(k_0,k,0,k_s) F_0(0,k_s)\right]
\cr
                    &-&    \frac{\pi}{2M} \int_0^{\infty} \frac{{k'}^2 dk'}{\varepsilon_{k'} (2\varepsilon_{k'}+M) }      W_0^S(k_0,k, a_+,k') F_0(a_+,k' )   \cr
                 &+&  \frac{i}{2M}  \int_0^{\infty}  \frac{{k'}^2dk' }{\varepsilon_{k'}}   \int_0^{\infty} dk'_0 \left[ \frac{ W^S_{0}(k_0,k,k'_0,k') F_0(k'_0,k')  - W^S_{0}(k_0,k,a_-,k')  F_0(|a_- |,k')}{  {k'}_0^2-a_-^2 }\right]  \cr
                &-&    \frac{i}{2M}  \int_0^{\infty} \frac{{k'}^2dk'} {\varepsilon_{k'}}     \int_0^{\infty} dk'_0 \left[ \frac{ W^S_{0}(k_0,k,k'_0,k') F_0(k'_0,k')  - W^S_{0}(k_0,k,a_+,k') F_0(a_+,k')}  {{k'}_0^2-a_+^2 } \right]
                   \label{Eq_F_sym}
\end{eqnarray}
\end{widetext}
Notice the appearance of the absolute value in the argument $| a_-|$ and  that the relation $W_0^S(k_0,k,0,k_s)=2W_0(k_0,k,0,k_s)$
accounts for the factor 8 in the denominator of the non-integral term.
We remind also that $M=2\varepsilon_{k_s}$.

The origin of the different terms appearing in (\ref{Eq_F_sym}) is quite clear:
\begin{itemize}
\item The non-integral term (second term in the first line)
is a sum of two equal contributions in (\ref{Integrand}): 
the first one comes from the two $\delta$-functions term  $I_2$, Eq. (\ref{2delta}), and 
the second one comes from the term without $\delta$-functions,  precisely the last term in the subtraction (\ref{subtr1a}) given by Eq. (\ref{nonint1}).

\item The one-dimensional integral terms (second and third lines)  result from the contribution $I_1$, Eq. (\ref{I1b}), i.e., from
the four contributions of one $\delta$-function terms -- $\delta(k'_0\pm a_-)$ in second line and $\delta(k'_0\pm a_+)$ in third line -- after integration over $k'_0$.  

\item The last two lines come from the product of four principal values  (no  $\delta$-function), the contribution $I_0$, Eq. (\ref{I0}), however without the term 
(\ref{nonint1}), which is incorporated in the non-integral term in the first line of the equation  (\ref{Eq_F_sym}).
\end{itemize}
The differences appearing in  squared brackets (2nd, 4th and 5th lines) correspond to the subtractions (\ref{subtr}) and (\ref{subtr1a}) used
to remove the pole singularities: $2\varepsilon_{k'}=M$  in second line, $k'_0=a_-$ in forth and $k'_0=a_+$ in fifth lines  respectively.    

\section{Kernel singularities}\label{kernel}

In view of the numerical integration of  (\ref{Eq_F_sym}), it is useful to know the precise positions of the singularities
both in the kernel and in the Born term.

The above considerations were devoted to treat  the poles of the free constituent propagators.  
However these singularities are not the only ones.  
The  propagator of the exchanged particle, i.e. the kernel (\ref{obe}), also has  two poles which, after  partial wave decomposition, turn into the logarithmic singularities of the $W_0$ kernel (\ref{W0}). 
Though the log-singularities can be  integrated numerically by "brut force", to improve precision, it is useful to treat them too. 
Their positions are found analytically, both in $k'_0$ and $k'$ variables. 

\bigskip
Let us first consider the singularities on $k'_0$.
It follows from   (\ref{W0}),  that $W_0$  is singular when \mbox{$\mid\eta(k_0,k,k'_0,k')\mid =1$},  where $\eta$ is defined by  (\ref{eta}).
Solving two equations $\eta=\pm 1$ relative to $k'_0$ we find the four singularities of $W_0$:
\begin{eqnarray}
k'_0&=&\phantom{-}k_0+\sqrt{\mu^2+(k\pm k')^2}, \cr
k'_0&=&\phantom{-}k_0-\sqrt{\mu^2+(k\pm k')^2},  \label{W0sing}  
\end{eqnarray}
The symmetric kernel $W_0^S$ has four additional singularities   when $\mid\eta(k_0,k,-k'_0,k')\mid= 1$. 
Together with (\ref{W0sing}), it means that $W_0^S$ is singular at the eight points:
\begin{eqnarray}
k'_0&=&\phantom{-}k_0+\sqrt{\mu^2+(k\pm k')^2}, \cr
k'_0&=&\phantom{-}k_0-\sqrt{\mu^2+(k\pm k')^2}, \cr
k'_0&=&-k_0+\sqrt{\mu^2+(k\pm k')^2}, \cr
k'_0&=&-k_0-\sqrt{\mu^2+(k\pm k')^2}  \label{W0Ssing}   
\end{eqnarray}
However, since the equation (\ref{Eq_F_sym})  was reduced to the interval $0<k'_0 <\infty$, 
one should take into account only the singularities on the positive axis $k'_0>0$. This is equivalent to take the absolute value of (\ref{W0Ssing}), that is
$W_S$ is singular at the four $k'_0$ values:
\begin{eqnarray*}
k'_0&=& \mid k_0+\sqrt{\mu^2+(k\pm k')^2} \mid \\
k'_0&=& \mid k_0-\sqrt{\mu^2+(k\pm k')^2}  \mid
\end{eqnarray*}
Their positions depend on the integration variable $k'$ (moving singularities) as well on the external momenta $k_0$ and $k$.

The numerical integration over $k'_0$  is split into as many intervals as needed in order to contain a single singularity in only one of its borders.  
The integral over each of these intervals is made safely by choosing an appropriate change of variable. 

\bigskip
The kernel singularities in the $k'$-variable manifest themselves only in the 2nd and 3rd lines of equation  (\ref{Eq_F_sym}).
They are also given by the  solutions,   with respect to $k'$, of  \mbox{$ \eta(k_0,k,k'_0,k')=\pm 1$} and  \mbox{$ \eta(k_0,k,-k'_0,k')= \pm 1$} for the particular values $k'_0= a_{\pm}$.
For the symmetrized $W_0^S$ kernel, this gives the  positions detailed in what follows.

The term with $k'_0=a_-$,  written in the second line of Eq. (\ref{Eq_F_sym}), is singular at:
\begin{equation*}
k'_-=    \frac{\epsilon kQ_+ \pm d_+ \sqrt{   Q_+^2-  4m^2(d_+^2-k^2) }Ã}{2(d_+^2-k^2)} ,  \quad \epsilon=\pm1
\end{equation*}
where
$$
 Q_+ =d_+^2-k^2 +m^2-\mu^2 ,         \quad  d_+ = k_0+{M\over2}, \quad M=2\varepsilon_{k_s}.
$$
That is, the integrand of the second term vs. $k'$ has the four  singularities denoted by  $k_{-,1,2,3,4}$:
\begin{eqnarray}
k'_{-,1}  &=& \frac{+kQ_+  + d_+ \sqrt{   Q_+^2-  4m^2(d_+^2-k^2) }Ã}{2(d_+^2-k^2)} , 
\nonumber\cr
k'_{-,2}  &=& \frac{+kQ_+   - d_+ \sqrt{    Q_+^2-  4m^2(d_+^2-k^2) }Ã}{2(d_+^2-k^2)} ,  
\nonumber\cr
k'_{-,3}  &=& \frac{-kQ_+   + d_+ \sqrt{   Q_+^2-  4m^2(d_+^2-k^2) }Ã}{2(d_+^2-k^2)} ,  
\nonumber\cr
k'_{-,4}  &=& \frac{-kQ_+   - d_+ \sqrt{    Q_+^2-  4m^2(d_+^2-k^2) }Ã}{2(d_+^2-k^2)} . 
\\
&&
\label{k-}
\end{eqnarray}

\bigskip
The term with $k'_0=a_+$, written in the third line of Eq. (\ref{Eq_F_sym}), is singular at:
\begin{equation*}
k'_+=  \frac{\epsilon kQ_-  \pm |d_-|\sqrt{   Q_- ^2-  4m^2(d_-^2-k^2) }Ã}{2(d_-^2-k^2)} ,   \qquad \epsilon=\pm1
\end{equation*}
where
\[  Q_- =d_-^2-k^2 +m^2-\mu^2,   \qquad d_-= k_0-{M\over2}  \]
That is the integrand has the four  singularities $k'_{+,1,2,3,4}$. Their positions are obtained from eqs. (\ref{k-}) by the replacement 
$Q_+\to Q_-$,  $d_+\to d_-$.

Since the integration domain of the $k'$ variable is positive, one should take into account only the real and positive values of the above singularities.

\subsection{Born term}

The S-wave Born term (\ref{FB}) is singular both  in $k_0$ and $k$ variables.
The singularities are logarithmic  in its real part and Heaviside-like discontinuities in the imaginary one.
Their positions can be found from the condition $\mid\eta(k_0,k,k'_0=0,k'=k_s)\mid=1$, where $\eta$ is defined in (\ref{eta}).

\begin{figure}[hbtp]
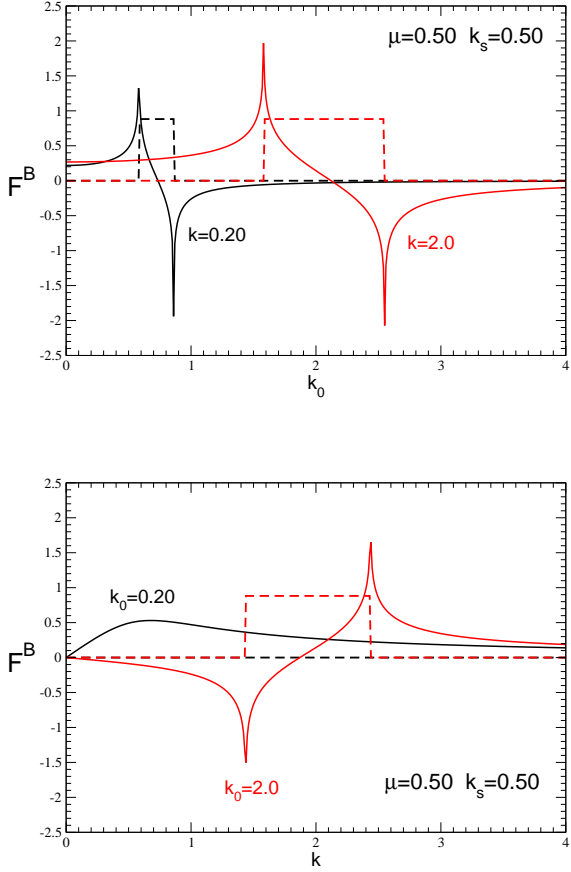
 
\begin{center}
\includegraphics[width=7.5cm]{FB_k0_1.123_0.50_0.50.eps}

\vspace{1cm}

\includegraphics[width=7.5cm]{FB_k_1.123_0.50_0.50.eps}
\end{center}
\caption{Singularities of the  S-wave Born term $F^B_0$ as a function of $k_0$ for two different values  of $k=0.20$ and $k=2.0$ (upper panel)
and as a function of $k$ for two different values of $k_0=0.20$ and $k_0=2.0$ (lower panel).}\label{FIG_FB_k0}
\end{figure}

In $k_0$-variable (see Fig. \ref{FIG_FB_k0} upper part), there are two singularities at the points
$$ k_0(k)=\sqrt{(k\pm k_s)^2+\mu^2} $$
for any value of $k$.

In the variable  $k$ (see Fig. \ref{FIG_FB_k0} bottom) the singularities (in its definition domain $k>0$) are given by the same equation rewritten in the form:
\[    k(k_0) = \left| k_s \pm  \sqrt{ k_0^2-\mu^2}  \right| \]
and exist only  for $k_0>\mu$.

\bigskip
The Minkowski BS amplitude $F_0$  has many non-analyticities due to its Born term $F_0^B$ and to the interaction kernel, including inelastic threshold effects. 
However the only singularities (infinite values and discontinuities) in the physical domain of its arguments are those originated by the Born term $F_0^B$ itself. 
Their existence makes difficult  representing $F_0$ on a basis of regular functions in view of a numerical solution of Eq. (\ref{BSE1}).
To circumvent this problem we factorize out the Born amplitude  by making the replacement
\begin{equation}\label{f0}
F_0=F_0^B \;\chi\; f_0 
\end{equation}
where $f_0$ is a smoother function  obeying the BS transformed equation  
\[   F_0=F_0^B+ K F_0   \quad \Rightarrow\quad f_0 =  {1\over\chi} + {1\over \chi F_0^B} K F_0^B \chi f_0   \]
$\chi$ is an arbitrary but suitable function introduced to provide a  convenient inhomogeneous term .
After that, the singularities of the $F_0$ are casted into the kernel and integrated using the same procedure as above.
We obtain in this way a non-singular equation for  a non singular function $f_0$ which can be solved by standard methods. 

The  off-mass shell BS amplitude $F_0$ in Minkowski space can be thus safely computed.

\section{Extracting scattering observables}\label{phase}

The amplitude $F(k,k_s;p)$ satisfying the BS equation (\ref{BSE}) is related to the $S$-matrix by:
\[ S=1+i(2\pi)^4\delta^{(4)}(p-p_f)F(k,k_s;p)\]
The unitarity condition for $S$-matrix $S^\dagger S=1$ 
is rewritten in terms of the amplitude $F^{on}(k,k_s;p)$ (which is on the  mass-shell $k_1^2=k_2^2=k_{1s}^2=k_{2s}^2=m^2$)  as follows:
\begin{widetext}
\begin{equation}\label{fpw2a}
i( {F^{on}}^\dagger -F^{on})=\int {F^{on}}^\dagger F^{on} (2\pi)^4\delta^{(4)}(p-k_1-k_2)\frac{d^3k_1}{(2\pi)^3 2\varepsilon_1}   \frac{d^3k_2}{(2\pi)^3 2\varepsilon_2} 
 \end{equation}
\end{widetext}
The sum over intermediate states in the product $S^\dagger S$ is understood as integration with the measure given in (\ref{fpw2a}). 
After substituting the partial waves decomposition (\ref{fpw1}) in the equation (\ref{fpw2a}) the latter obtains the form:
\begin{equation}\label{fpw3}
i({F^{on}}^*_L-F^{on}_L)=\frac{2k_s}{\varepsilon_{k_s}}\left| F^{on}_L\right|^2
 \end{equation}
where  \mbox{$F^{on}_L\equiv F_L(k_0=0,k=k_s)$} is the on-shell amplitude.
The function satisfying Eq. (\ref{fpw3}) is represented as:
\begin{equation}\label{fpw4}
F^{on}_L=\frac{\varepsilon_{k_s}}{k_s}\exp(i\delta_l)\sin \delta_l
 \end{equation}
with arbitrary real $\delta_l$.   Solving Eq. (\ref{fpw4}) relative to $\delta_l$ we find that 
the on-shell amplitude determines the phase shift according to:
\begin{equation}\label{delta}
\delta_L(k_s)=\frac{1}{2i}\log\Bigl(1+\frac{2i k_s } {\varepsilon_{k_s}} F^{on}_{L}\Bigr)
\end{equation}

\section{Euclidean scattering amplitude}\label{euclid_scat}

Equation (\ref{Eq_F_sym}), is free from the poles of the constituent propagators 
and after the appropriate treatment of the logarithmic singularities provides the desired solution for the off-shell amplitude in Minkowski space. 
This equation is however rather cumbersome, even in the simplest case of two scalar particles in S-wave we are considering,
and looks rather different from the initial BS equation (\ref{BSE}). 
It would thus be of the highest interest to have at our disposal an independent test of the numerical solutions.

\begin{figure}[hbtp] 
\centering
\includegraphics[width=7.5cm]{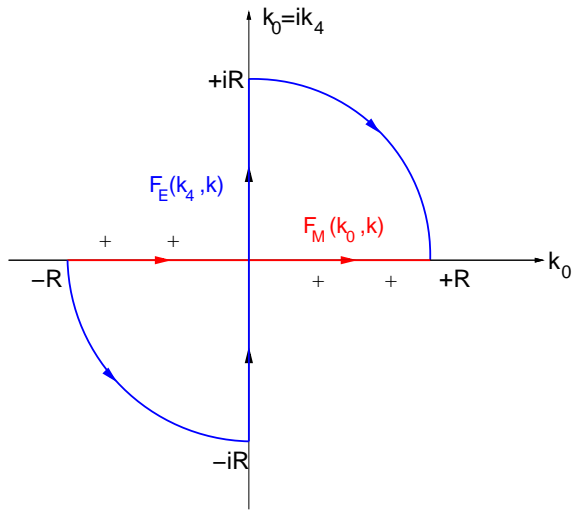}
\caption{Minkowski ($F_M$) and Euclidean ($F_E$) BS amplitudes   in the complex $k_0$ plane with (in crosses) bound state singularities.}\label{FIG_F_M_F_E}
\end{figure}

The test we have performed is based on an Euclidean  version of the initial 
BS equation for the scattering amplitude in Minkowski space (\ref{BSE}),  formally writen as
\begin{equation}\label{EQM}
F^M = F^{M,B} + K^MF^M 
\end{equation}
We can define the so called Euclidean BS scattering amplitude (see Fig. \ref{FIG_F_M_F_E}) by
\[  F^E(k_4,k)  =F^M(k_0=ik_4,k)   \]
By applying the Wick rotation $k_0=ik_4$ to (\ref{BSE}) we will derive in what follows the equation satisfied by $F^E$.
We will introduce all along this section the index M or E to distinguish between both amplitudes quantities.  

\bigskip
Some preliminary remarks are in order:
\begin{enumerate}

\item  Contrary to the Minkowski case, the off-shell Euclidean amplitude  cannot be used to compute physical observables like e.m.  form factors
even in the bound state case. The reason is the impossibility to make the Wick
rotation in the form factor integral \cite{ckm_ejpa,ck-trento}.

\item The on-mass-shell condition for $k_0=0$ corresponds to $k_4=0$. 
Therefore both amplitudes, although obeying different equations, should coincide on the mass shell:  \mbox{$F^M(k_0=0,k_s)=F^E(k_4=0,k_s)$} 
and should thus provide the same phase shifts. 
This property will be used to check our Minkowski results.

\item  In the case of the scattering states, the Wick rotation  cannot be performed in a naive way to  the equation (\ref{EQM})
by simply replacing $k_0\to k_4=-ik_0$. This important point will be developed below.
\end{enumerate}

\begin{figure}[htbp]
\centering\includegraphics[width=7.5cm]{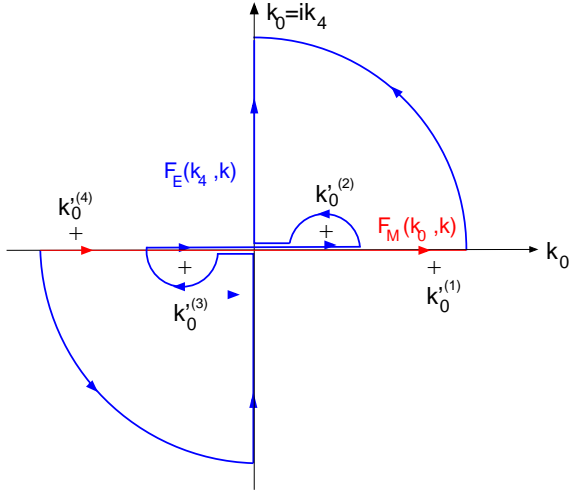}
\caption{Singularities of the propagators for scattering state and the integration contour after rotation in the complex plane $k_0$,
if $k'<k_s$.}\label{fig1a}
\end{figure}

\subsection{Rotating the integration contour}\label{rot}

Let us consider the pole positions appearing in the equation (\ref{BSE1}) and given by (\ref{Poles}).
The integration domain  $k'<k_s$ corresponds to $\varepsilon_{k'}<\varepsilon_{k_s}$ and hence to 
$Re[{k'}^{(2)}_0]>0$ and $Re[{k'}^{(3)}_0]<0$.
The positions of these singularities are illustrated in Fig. \ref{fig1a}. 
The contour cannot be anti-clock-wise rotated without taking into account the residues at these poles  ${k'}^{(2)}_0,{k'}^{(3)}_0$.
Notice that for the bound state problem, the value $\varepsilon_{k_s}$ is replaced by $\frac{M}{2}$.
Since  for any $k'$ we have $\varepsilon_{k'}>\frac{M}{2}$, when performing the Wick rotation  in the bound state equation 
there is no any crossed singularity and so no any additional contribution.

Therefore, for the bound state case, the positions of singularities allows to safely rotate the contour as it is shown in Fig.  \ref{FIG_F_M_F_E}.
In this way, one gets equation for the Euclidean BS amplitude $F^E$.

On the contrary, for the scattering state case, 
the rotated  contour in the complex plane of $k'_0$  (see Fig. \ref{fig1a}) 
crosses  two of the four pole singularities ${k'}^{(2)}_0,{k'}^{(3)}_0$ displayed in the integrand of Eq. (\ref{BSE}). 
They are on the real axes $k'_0$ at
 the points $k'_0=\pm (\varepsilon_{k_s}-\varepsilon_{k'})$. 
 The residues in these poles (to be added to the integral terms) are still expressed  in terms of the Minkowski off-shell amplitude: 
\begin{equation}\label{EMink}
 \tilde{F}^M_L(k')\equiv F^M_L(k'_0= \varepsilon_{k_s}-\varepsilon_{k'},k'). 
\end{equation}
Thus, when transforming Eq. (\ref{EQM}) into Euclidean,  we obtain  not an Euclidean equation, but a  non-singular equation
which indeed contains in l.h.-side the Euclidean BS amplitude $F^E$, however, in r.h.-side, under the integrals, both Euclidean amplitude $F^E$ and  
the particular Minkowski amplitude  $\tilde{F}^M_L(k)$ defined in (\ref{EMink}). 
One can similarly derive another non-singular equation which  contains in l.h.-side the particular Minkowski amplitude  $\tilde{F}^M_L(k)$, and in r.h.-side, under the integrals, again both Euclidean amplitude $F^E$ and  the particular Minkowski amplitude  $\tilde{F}^M_L(k)$. 
In this way, we derive the system of two equations 
which couples the Euclidean amplitude  $F^E_L(k_4,k)$ to  the particular Minkowski off-shell  amplitude  $\tilde{F}^M_L(k)$. 
A similar derivation is described in \cite{tjon}. 

An additional test is to check that the off-shell Minkowski amplitude $F_L^M(k_0,k)$
obtained by solving  equation (\ref{BSE1}) coincides, for the particular value $k_0=\pm (\varepsilon_{k_s}-\varepsilon_{k})$, with the independent solution 
$\tilde{F}^M_L(k)$ of the system of equations. Furthermore, we can check that all three on-mass amplitudes (for the S-waves, in particular) coincide with each other:
$$F_0^M(k_0=0,k=k_s)=F^E_0(k_4=0,k=k_s)=\tilde{F}^M_0(k_s)$$
and therefore give the same phase shifts.

Below in this section we sketch the derivation of this system of two equations which couples the amplitudes  $F^E(k_4,k,z)$ and $\tilde{F}^M(k,z)$. In this derivation we will consider  the case when the amplitudes are not decomposed in the partial waves. The two equations for the \mbox{S-wave} amplitudes  $F^E_0(k_4,k)$ and 
$\tilde{F}^M_0(k)$ solved numerically will be given in appendix  \ref{first}.

We should also analyze the position of singularities in the kernel (\ref{obe}). They are at the points:
\[ {k'}^{\pm}_0=k_0\pm\sqrt{(\vec{k}-\vec{k'})^2+\mu^2}\mp i\epsilon\]
When we rotate the contour around  $k'_0=0$ we must distinguish two cases.
 If $|k_0|< \sqrt{(\vec{k}-\vec{k'})^2+\mu^2}$, the singularity
\begin{equation}\label{kp0+}
{k'}^+_0=k_0+\sqrt{(\vec{k}-\vec{k'})^2+\mu^2}- i\epsilon
\end{equation}
 is in the 4th quadrant and  
\begin{equation}\label{kp0-}
{k'}^-_0=k_0-\sqrt{(\vec{k}-\vec{k'})^2+\mu^2}+ i\epsilon
\end{equation}
 in the 2nd one.  
In this case the contour can be rotated anti-clockwise without crossing the singularities. 
If $|k_0|> \sqrt{(\vec{k}-\vec{k'})^2+\mu^2}$,  both
singularities are in the same half-plane (e.g., at the right
half-plane, if $k_0>0$) and  the contour cannot be rotated.

However, Wick rotation must be performed in both $k'_0$ and $k_0$ variables simultaneously. 
That is, {rotating}  the integration contour in  $k'_0$ by an angle $\phi$, we change also the  variable
$k_0\to k_0\exp(i\phi)$. Then the positions of singularities (\ref{kp0+}) and (\ref{kp0-}) are also rotated.
As it can be easily checked, ${k'}^-_0$ is rotated faster than
the contour and ${k'}^+_0$ is rotated slower; therefore they move away
from the contour and the contour rotation can be safely done.
Its  final position, after rotation by
$\phi=\pi/2$, is $-i\infty\ < k'_0 < i\infty$, whereas the final value of $k_0$ turns into $ik_0$. 
The singularities of the propagator (\ref{obe}) do not prevent from the Wick rotation in both variables.

For the amplitude
$F(\vec{k},k_0=\varepsilon_{k_s}-\varepsilon_{k};\vec{k_s})$
the condition
$|k_0|< \sqrt{(\vec{k}-\vec{k'})^2+\mu^2}$ turns into
$$
\varepsilon_{k_s}-\varepsilon_{k'}<\sqrt{(\vec{k}-\vec{k'})^2+\mu^2}
$$
Since we integrate over $\vec{k'}$, the maximal value of l.h.-side is $\varepsilon_{k_s}-m$ and the minimal value of r.h.-side is $\mu$. 
Hence, this inequality is violated if  $\varepsilon_{k_s}-m>\mu$, that is when $\varepsilon_{k_s}>m+\mu$ $\rightarrow$ $\sqrt{s}>2m+2\mu$, 
i.e. above the two-meson creation threshold.
The singularities of the one-boson exchange kernel allow the contour
rotation only in this kinematical domain. Above that,
additional contributions  should be taken into account. 
The same conclusion was found in \cite{tjon}. 
In our solution  of the Euclidean equation, we will not exceed the two-meson creation threshold.

\subsection{Euclidean equation}\label{Euclidean_equation}

We start with performing the Wick rotation shown in Fig. \ref{fig1a} to  the equation (\ref{BSE}).
In the c.m.-frame $\vec{p}=0$,  it is transformed into:
\begin{widetext}
\begin{equation}\label{BSEc}
F^E(k_4,\vec{k};\vec{k}_s)=V^B(k_4,\vec{k};\vec{k}_s)+ \int\frac{d^4k'}{(2\pi)^4}
\frac{V(k_4,\vec{k'};k'_4,\vec{k'}) F^E(k'_4,\vec{k'};\vec{k}_s)}
{({k'_4}^2 +a_-^2)({k'_4}^2+a_+^2)} + {S}
\end{equation}
\end{widetext}
where 
\begin{equation}\label{V}
V(k_4,\vec{k};k_4',\vec{k'})=\frac{16\pi m^2\alpha}{(k_4-k'_4)^2+(\vec{k}-\vec{k'})^2+\mu^2},
\end{equation}
\begin{equation}\label{Vinh}
V^B(k_4,\vec{k};\vec{k}_s)=V(k_4,\vec{k};k_4'=0,\vec{k'}=\vec{k}_s)
\end{equation}
and $S$ denotes the contribution due to the two singularities shown in Fig. \ref{fig1a}. 
This contribution, existing only if $\varepsilon_{k_s} -\varepsilon_{k'}>0$,  is given by the sum of two residues $S=S_1+S_2$;
 $S_1$ is the contribution from  $k'_0=(\varepsilon_{k_s} -\varepsilon_{k'})+i\epsilon$  multiplied by ($2\pi i$)
and $S_2$ the one from $k'_0=-(\varepsilon_{k_s} -\varepsilon_{k'})-i\epsilon$ multiplied by (-$2\pi i$).

Contribution $S_1$ has the form:
\begin{widetext}
\begin{equation}\label{sing1}
S_1(k_0)=\frac{\pi g^2}{4(2\pi)^4} \frac{\tilde{F}^M(k',z')}{\varepsilon_{k_s} \varepsilon_{k'}[-a_- +i\epsilon]
\left[-a_- + \sqrt{(\vec{k'}-\vec{k})^2+\mu^2} -k_0\right]  \left[-a_- -  \sqrt{(\vec{k'}-\vec{k})^2+\mu^2} -k_0 +i\epsilon\right]}  
\end{equation}
\end{widetext}
whereas $S_2$  is given by  $S_2(k_0)=S_1(-k_0)$. The sum $S_1+S_2$ is symmetric relative to $k_0\to - k_0$, as should be. 
At this point it is interesting to keep these expressions for $S_{1,2}$  with $k_0$ not replaced by $ik_4$.
The reason will become clear later, when,  deriving another equation, we will substitute \mbox{$k_0=\varepsilon_{k_s}-\varepsilon_{k}$.}
Above the $2m+\mu$ inelastic threshold these factors give  an imaginary contribution making the elastic phase shift
complex.  The above form of $S_{1,2}$ is convenient to find this imaginary part. 

Setting $k_0=ik_4$ the preceding expression reads:
\begin{widetext}
\begin{small}
\begin{equation}\label{sing}
S(ik_4)=\frac{g^2 \pi}{(2\pi)^4}  \int_{k'<k_s}d^3k'\frac{\tilde{F}^M(k',z')} {2\varepsilon_{k'} \varepsilon_{k_s}( a_- -i\epsilon)}   
\frac{\Bigl[k_4^2-\Bigl(a_- -\sqrt{(\vec{k'}-\vec{k})^2+\mu^2}\,\Bigr)
\Bigl(a_- +\sqrt{(\vec{k'}-\vec{k})^2+\mu^2}\,\Bigr)\Bigr]}
{\Bigl[k_4^2+\Bigl(a_- -  \sqrt{(\vec{k'}-\vec{k})^2+\mu^2}\,\Bigr)^2\Bigr]
\Bigl[k_4^2+\Bigl(a_- +\sqrt{(\vec{k'}-\vec{k})^2+\mu^2}\,\Bigr)^2\Bigr]}   
\end{equation}
\end{small}
\end{widetext}
Equation (\ref{BSEc}) is not singular.
The factor \mbox{$1/[{k'_4}^2 +a_-^2(k')]$} in the integrand of (\ref{BSEc}) is singular when 
$k'_4=0$ and $k'=k_s$ simultaneously, but this singularity is in fact canceled by a similar  term in $S$ given by (\ref{sing}).
It is however convenient to cancel these two singularities explicitly and analytically and obtain a regular resulting expression. 
The transformations are elementary but lengthy and will not be carried out in detail.
Below  are indicated the main steps.

We make subtraction in the integrand and add the subtracted term:
\begin{widetext}
\begin{eqnarray}
F^E(k,k_4,z)&=& V^{B}(k_4,\vec{k},\vec{k_s})
+  \frac{1}{(2\pi)^4}
\int d^3k'\; \int_{-\infty}^{\infty}dk'_4 V(k_4,\vec{k};k'_4,\vec{k'})  
\left\{
\frac{F^E({k'}_4,k',z')}{({k'_4}^2 +a_-^2)({k'_4}^2+a_+^2)}
-
\frac{F^E(0,k',z')}
{({k'_4}^2 +a_-^2)
a_+^2}\right\}  \cr
&+&\frac{1}{(2\pi)^4}
\int d^3k'\; \int_{-\infty}^{\infty}d{k'}_4
\frac{V(k_4,\vec{k};k'_4,\vec{k'})F^E(0,k',z')}
{({k'_4}^2 +a_-^2)a_+^2}   + S[ \tilde{F}^M  ]
\label{eq15}
\end{eqnarray}
\end{widetext}
There is no any singularity in the difference. The  subtracted term is not unique. 
Our choice was motivated in order to obtain an analytic result for the integral over $dk'_4$ in the last line of Eq. (\ref{eq15}).

The  term which we add is singular  
$k'=k_s$.  We perform an additional subtraction  to eliminate this singularity and again add the subtracted term. 
This additional term is, of course, again singular but analytic. 
Its contribution, after integration in the limits $ 0<k'<k_s-\delta, \quad k_s+\delta < k' <\infty$ and at  $\delta \to 0 $,
is $\sim \log(\delta/m)$. It is exactly cancelled analytically by a  similar term in the singular part of $S$. 
The resulting S-wave equation is regular and given in appendix \ref{first}, Eq. (\ref{eq18}). 

The equation (\ref{eq15}) and, equivalently (after cancellation of singularities), Eq. (\ref{eq18})  relate the Euclidean amplitude $F^E(k_4,k,z)$ and the Minkowski one $\tilde{F}^M(k,z)$ appearing in $S$.
To determine both amplitudes, we should obtain an additional equation.

This new equation is still obtained by performing a  Wick rotation $k'_0=ik'_4$ to  (\ref{BSE}).
However, instead of taking $k_0=ik_4$ we set $k_0=\varepsilon_{k_s}-\varepsilon_{k}$ for $k<k_s$. 
As discussed at the end of the subsection \ref{rot}, for this particular value of $k_0$ the kernel singularities do not prevent from the Wick rotation below the two meson creation threshold. In this way we get the following equation (symbolically):
\begin{equation}\label{tildF}
 \tilde{F}^M_0(k,z)=\mbox{r.h.-side Eq.(\ref{eq15})  at }[k_4=i(\varepsilon_{k_s}-\varepsilon_{k})],
 \end{equation}
that is  the right-hand side term of the equation (\ref{eq15}) taken at the value $k_4=i(\varepsilon_{k_s}-\varepsilon_{k})$. 
The corresponding explicit S-wave equation  is  given in appendix \ref{first}, Eq. (\ref{eq18b}).

\section{Results}\label{num}

We present in this section the results of solving the BS equation  in Minkowski space (\ref{Eq_F_sym})
and the coupled Euclidean-Minkowski system of equations  (\ref{eq18}) and (\ref{eq18b}).
Some details of the numerical methods used are given in the Appendix  A.

We have first computed  the bound state solutions, denoted $\Gamma_0(k_0,k)$, by dropping the inhomogeneous term $F_0^B(k_0,k)$ in (\ref{Eq_F_sym}) and setting $M=2m-B$.
The binding energies $B$  thus obtained  coincides, within four-digit accuracy, with the ones calculated in our previous work \cite{bs1}.

\vspace{1.0cm}
\begin{figure}[hbtp]
\centering
\includegraphics[width=7.5cm]{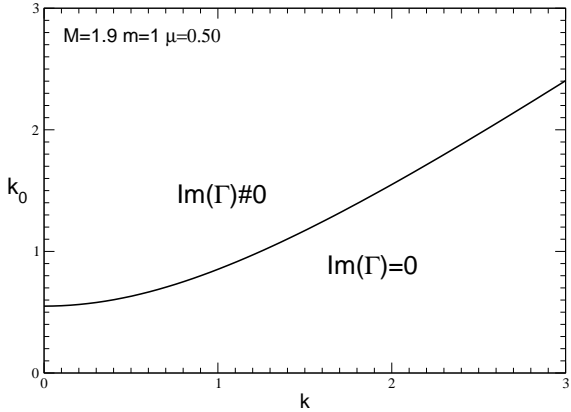}
\caption{Locus for Im$(\Gamma)=0$ in the ($k_0,k$) plane for $m=1$ $M=1.90$ and $\mu=0.5$.
Below this curve, Im$(\Gamma)=0$ and above  Im$(\Gamma)\neq0$.}\label{FIG_locus}
\vspace{1.5cm}
\end{figure} 
An interesting issue is the appearance  of an imaginary part in $\Gamma_0(k_0,k)$
when normalized for instance by $\Gamma(0,0)=1$.
Inspite of this real normalization condition, $\Gamma_0(k_0,k)$ will become imaginary
depending on the kinematical domain of its arguments corresponding to the virtual meson creation.
Written in terms of variables $k_i$, defined in (\ref{k_i}), they read:
\begin{equation}
k_1^2  >(m+\mu)^2 \quad {\rm or}\quad
k_2^2 > (m+\mu)^2 
\end{equation}
which in terms of variables $(k_0,k)$ in the center of mass frame, become:
\begin{eqnarray*}
\left( {M\over2} + k_0\right)^2 - \vec{k}^2   &>&(m+\mu)^2 \cr
\left( {M\over2} - k_0 \right)^2 - \vec{k}^2   &>& (m+\mu)^2 
\end{eqnarray*}
$\Gamma$ obtains an imaginary part if one of these two preceding conditions is fulfilled.
In the ($k_0,k$) plane and for positive values of $k_0$ this gives the locus
\begin{equation}\label{k0_k}
k_0(k) =  - {M\over2}  + \sqrt{k^2  + (m+\mu)^2}
\end{equation}
Above this curve (represented in Fig. \ref{FIG_locus}) the imaginary part of $\Gamma\neq0$.

\begin{figure}[hbtp] 
\centering
\includegraphics[width=7.5cm]{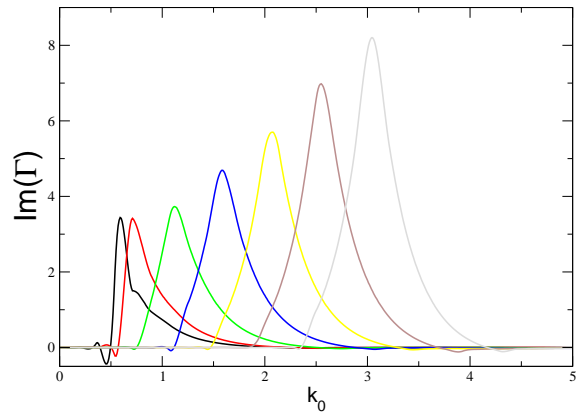}
\caption{Im$[\Gamma(k_0,k)]$ as a function of $k_0$ for different values of $k$. It corresponds to  $\alpha=1.44$, $B=0.01$ and $\mu=0.5$.}\label{FIG_Im_Gamma_k0_28}
\end{figure}

We display in Fig. \ref{FIG_Im_Gamma_k0_28} the $k_0$-dependence of the imaginary part of amplitude $\Gamma(k_0,k)$ obtained in our calculations for different values of $k$. It corresponds to the parameters $\alpha=1.44$, $\mu=0.50$, $B=0.01$.
We can thus check that the a vanishing imaginary part of $\Gamma$  appears at the $k_0$ values given by equation (\ref{k0_k}).
 Note also the difficulty in reproducing a sharp non analytic threshold behavior in terms of smooth functions even if they
 are as flexible as splines. The small oscillations in the vicinity of the threshold
 are artefacts of our spline basis. They can be reduced by increasing the number of basis elements.
 
 \vspace{1.0cm}

The  scattering amplitude $F_0(k_0,k)$ in Minkowski  space has been  calculated  for $L=0$ states and the corresponding phase shifts have been 
extracted according to (\ref{delta}).

The BS relativistic formalism accounts naturally for  the meson creation in the scattering process, when the available  kinetic energy allows it.
The inelasticity threshold corresponding to n-particle creation is given by
\begin{equation}
k^{(n)}_s = m \sqrt{  \left( {\mu\over m}\right)  \;n  +    {1\over4}\left( {\mu\over m}\right) ^2 \; n^2 }
\end{equation}
Below the first inelastic threshold,  $k^{(1)}_s=\sqrt{m\mu+\mu^2/4}$, the phase shifts are real.
This unitarity condition is not automatically fulfilled in our approach, but appears as a consequence of handling the correct solution
and provides a stringent test of the numerical method.
Above  $k^{(1)}_s$, the phase shift obtains an imaginary part which behaves like
\begin{equation}\label{Imdel}
 {\rm Im}[\delta_0]\sim (k_s-k^{(1)}_s)^2
 \end{equation}
in the threshold vicinity.
Higher inelasticity thresholds, corresponding to creation of 2, 3, etc. intermediate mesons at $k^{(n)}_s$, are also  taken into account in our calculations.

\begin{widetext}
\begin{table}[hbtp]
\caption{\mbox{Real and imaginary parts of the phase shift (degrees) calculated by solving} \mbox{the  equation (\ref{Eq_F_sym}) vs. incident momentum $k_s$ for $\alpha=1.2$  and $\mu=0.5$. Corresponding}  \mbox{first inelastic threshold is $k^{(1)}_s=0.75$.}}\label{tab2}
\begin{center}
\begin{tabular}{lllll ll   l l l  l  l    l  l l }
\hline
$k_s$            & 0.05  & 0.1      &  0.2   &  0.3     & 0.4     &   0.5   &   0.6     &  0.7    & 0.8      &  0.9         & 1.0       & 1.3      & 1.5      &  2.0 \phantom{0} \;\;\;2.5 
 \phantom{0} \;\;3.0\\ \hline
$Re[\delta_0]$  & 124  &  99.9   & 77.8 &   65.1  & 56.2  &  49.3  &  43.9   &  39.4  & 35.7    &   32.5     & 29.7    &  22.8   &  19.3   &13.3\phantom{0} \; 9.78
 \; \;7.78 \\
$Im[\delta_0]$  & 0        &      0     &  0      &     0     & 0       &      0    &     0      &      0    & 0.033  &   0.221  & 0.453  & 0.848  & 0.852   & 0.578 \; 0.333 
\,  0.203  \\ \hline
\end{tabular}
\end{center}
\end{table}
\end{widetext}

Some selected results corresponding to  $\alpha=1.2$  and $\mu=0.5$ are listed in Table \ref{tab2}.
We have also solved the system of equations, derived in Sec.  \ref{euclid_scat} and Appendix C (eqs. \ref{eq18} and \ref{eq18b}),
 coupling the Euclidean amplitude to the  Minkowski one at the particular value  $k_0=\varepsilon_{k_s}-\varepsilon_{k}  $.
The phase shifts  found by these two independent methods are consistent to each other within the accuracy given in this table. 
They are also rather close to ones found in \cite{tjon}.

\begin{figure}[hbtp]
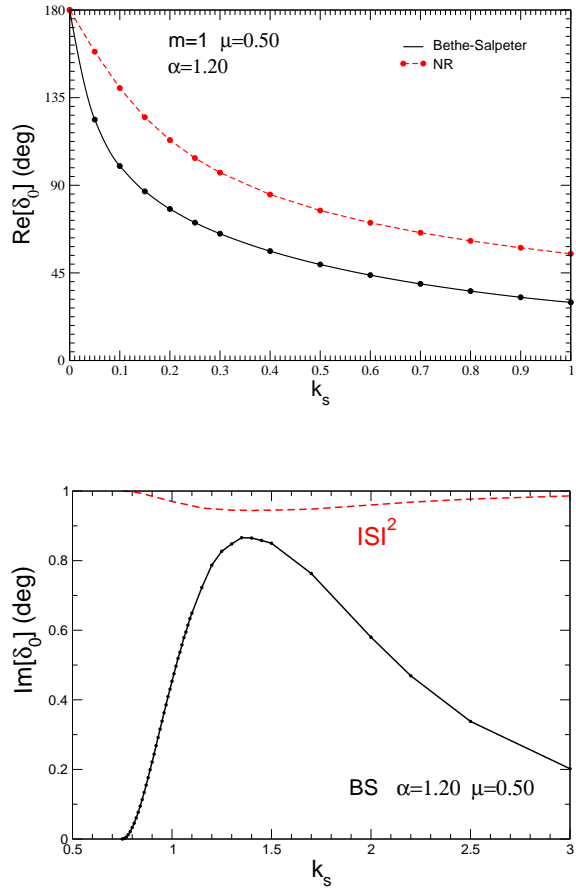
 
\begin{center}
\includegraphics[width=7.5cm]{Mod_Phaseshifts_alpha_1.200_0.50.eps} 

\vspace{1cm}

\includegraphics[width=7.5cm]{Long_Inel_alpha_1.20_mu_0.50.eps}
\end{center}
\caption{Upper panel: real  phase shift (degrees) for $\alpha=1.2$ and $\mu=0.50$ calculated  via BS equation (solid)  compared to the non-relativistic results (dashed). Lower panel: imaginary phase shift (degrees) calculated  via BS equation (solid) and the value $|S|^2=\exp(-2\mbox{Im}(\delta))$ (dashed).
\vspace{0.5cm}}\label{fig2}
\end{figure}

Figure \ref{fig2}  (upper panel) shows the real  phase  shifts calculated with BS  (solid line) as a function of the scattering momentum $k_s$.
They are compared to the non-relativistic (NR)  values (dashed lines) provided by the Schr\"odinger equation with the Yukawa potential.
For this value of $\alpha$ there exists a bound state and, according to the Levinson theorem, the phase shift starts at 180$^\circ$.
One can see that the difference between relativistic and non-relativistic results is considerable even for relatively small incident momentum.

The lower panel  shows the imaginary part of the phase shift. It appears  starting from the first inelastic meson-production threshold $k^{(1)}_s=0.75$ and displays the expected quadratic behavior (\ref{Imdel}).
Simultaneously the modulus squared of the S-matrix (dashed line) starts differing from unity.
The results of this figure contain  the contributions of the second $k^{(2)}_s=1.118$ meson creation threshold, the third  one $k^{(3)}_s=1.435$, etc., up to eight meson creation  threshold $k^{(8)}_s=2.828$.
Notice that, as mentioned in Sec. \ref{Euclidean_equation}, the applicability of the method coupling Euclidean and Minkowski amplitudes is limited in its present formulation
to the second inelastic threshold while our direct Minkowski space approach can go through.

\vspace{1.cm}
\begin{figure}[hbtp]
\centering
\includegraphics[width=7.5cm]{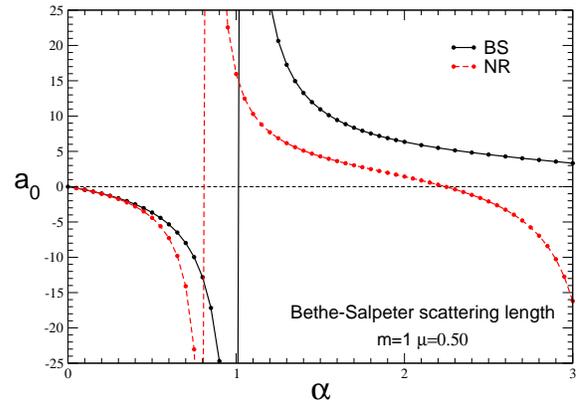}
\caption{BS scattering length $a_0$ versus the coupling constant  $\alpha$ (solid), compared to the non-relativistic results (dashed) for $\mu=0.5$.}\label{fig_a0}
\end{figure}
The low energy parameters were computed directly at $k_s=0$
and found to be consistent with a quadratic fit to the effective range function \mbox{$k\cot\delta_0(k)= -\frac{1}{a_0} + \frac{1}{2} r_0 k^2 $.}
The BS scattering length  $a_0$ as a function of the coupling constant $\alpha$ is given in Fig. \ref{fig_a0} for  $\mu=0.50$.
It is compared to the non-relativistic (NR)  values.
The singularities correspond to the appearance of the first bound state at $\alpha_0=1.02$ for BS and  $\alpha_0=0.840$ for NR.
One can see that the differences between a relativistic and a non-relativistic treatments of the same problem
are not of kinematical origin since even for  processes involving zero energy they  can be substantially large, especially in presence of bound state.
It is worth noticing that only  in  the limit $\alpha\to0$  the two curves are tangent to each other
and in this region the results are given by the  Born approximation
\begin{equation}
a^B_0=   - \; {1\over\mu} {m\over\mu} \alpha
\end{equation}
which is the same for the NR and the BS equation.
Beyond this region both dynamics are not compatible.
This non matching between NR and relativistic equations was already pointed out in \cite{MC_PLB_00,CKF_FBS54_2013} when computing  the binding energies $B$ of a two scalar and two fermion
system  in the limit $B\to0$ with different relativistic approaches. A recent work \cite{HL_PRD85_2012} devoted to this problem proposes the construction of equivalent  
non relativistic potential using the technique of  geometrical spectral inversion. It would be interesting to check the robustness of this equivalence by extend it to the scattering states and to form factors.

Some   numerical values of the scattering length for different values of  the exchanged mass $\mu$ were given in Ref.  \cite{CK_PLB_2013}, Table I.
It can be checked by direct inspection, that the scaling properties of the non relativistic equation \cite{SCQ_EPJA47_2011}, in particular  the relation between the
scattering length corresponding to different values of $\mu$ and coupling constants
\begin{equation}
 a_0\left( {\mu\over m} ,\alpha\right) = {1\over \mu} a_0\left(1,{\alpha\over {\mu\over m}} \right)
\end{equation}
are no longer valid except in the Born approximation region.

\begin{figure}[hbtp]
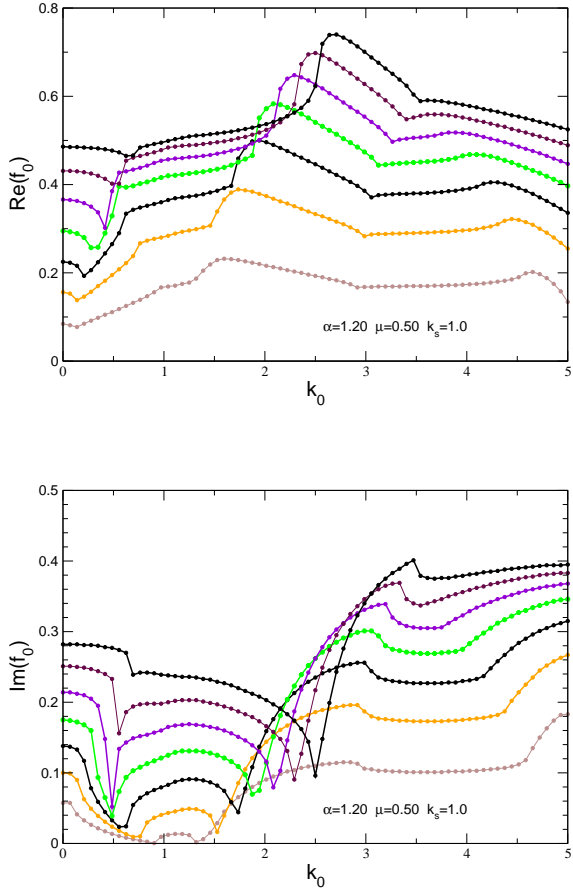
 
\includegraphics[width=7.5cm]{Ref_k0_F3.eps}

\vspace{1cm}

\includegraphics[width=7.5cm]{Imf_k0_F3.eps}
\caption{Real (upper panel) and imaginary (lower panel) parts of the factorized off-shell scattering amplitude $f_0$  defined in Eq.  (\ref{f0})
for $\alpha=1.2$ and $\mu=0.50$ vs. $k_0$ for different fixed values of $k$.}\label{FIG_f_k0}
\end{figure}

We display in Fig. \ref{FIG_f_k0} the factorized part of the off-shell scattering amplitude $f_0$, defined in (\ref{f0}),
as a function $k_0$ for different values of $k$.  It corresponds to the parameters $\alpha=1.2$, $\mu=0.5$ and $k_s=1.0$.
The upper panel displays its real part  and lower panel the imaginary one.
Its $k$-dependence is shown in Fig. \ref{FIG_f_k}  for different values of $k_0$.
The regular amplitudes $f_0$  are those effectively computed in our approach.
As one can see, these functions are no longer singular, although they present several sharp structures and cusps both on its real and imaginary parts.
The corresponding three-dimensional plots $F_0(k_0,k)$ for the values $\alpha=0.5$, $\mu=0.5$ and $k_s=0.5$ are given
in  Fig. 4 of \cite{CK_PLB_2013}. 
\vspace{0.5cm}

\begin{figure}[hbtp] 
\includegraphics[width=7.5cm]{Ref_k_F3.eps}

 \vspace{1cm}

\includegraphics[width=7.5cm]{Imf_k_F3.eps}
\caption{Real (upper panel) and imaginary (lower panel) parts of the factorized off-shell scattering amplitude $f_0$  defined in Eq.  (\ref{f0})
for $\alpha=1.2$ and $\mu=0.50$ vs. $k$ for different fixed values of $k_0$.\vspace{0.5cm}}\label{FIG_f_k}
\end{figure}
\bigskip

\begin{figure}[hbtp]
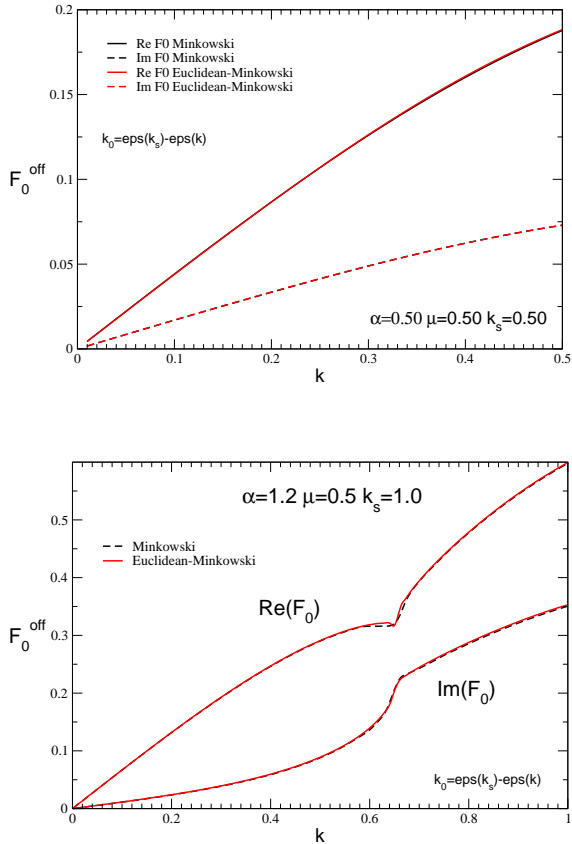

\includegraphics[width=7.5cm]{Comp_F0_off_0.50_0.50_0.50.eps}

\vspace{1cm}

\includegraphics[width=7.5cm]{Comp_F0_off_1.20_0.50_1.00.eps}
\caption{Comparison between the Minkowski amplitude  ${F_0}$ (in black) and   $\tilde{F_0}$ (in red) obtained by solving
the Euclidean-Minkowski coupled equations for the particular off shell value $k_0=\varepsilon_{k_s}-\varepsilon_{k} $  (see Sec. \ref{euclid_scat}).
Solid line denotes the real parts  of the amplitudes  and dashed line denotes their imaginary part.
Upper pannel corresponds to  $\alpha=0.5$, $\mu=0.50$ and $k_s=0.5$
and lower pannel  corresponds to  $\alpha=1.2$, $\mu=0.50$ and $k_s=1.0$.}\label{Foff_ks_1.0}
\end{figure}

The off-shell scattering amplitudes $F_0$ computed using our direct method in Minkwoski space have been tested using also the results of the coupled Euclidean-Minkowski equation derived in Sec. \ref{euclid_scat}.
As it was explained in Sec. \ref{euclid_scat}, these equations couple the Euclidean amplitude to the  Minkowski  one for the particular off-shell value  $k_0=\varepsilon_{k_s}-\varepsilon_{k} $, denoted $\tilde{F_0}(k)$.
The comparison of the off-shell amplitude $\tilde{F_0}(k)$ found by this method  with the solution of Eq. (\ref{Eq_F_sym}) $F_0(k)=F^M(k_0=\varepsilon_{k_s}-\varepsilon_{k},k)$ for the same argument $k_0$  provides an independent test of our direct method based on Eq. (\ref{Eq_F_sym}).

Two illustrative examples are shown  in  Fig. \ref{Foff_ks_1.0} for different values of the parameters $\alpha,\mu$ and $k_s$. 
Solid line denotes the real parts  of the amplitudes ${F_0}(k)$ (in black) and   $\tilde{F_0}(k)$ (red color), whereas the dashed line denotes their imaginary part.

Upper panel corresponds to  $\alpha=0.5$, $\mu=0.50$ and $k_s=0.5$. The results of  ${F_0}(k)$ and   $\tilde{F_0}(k)$
are not distinguishable by eyes, their relative difference being at the level of $10^{-3}-10^{-4}$.

Lower panel  corresponds to the parameters $\alpha=1.2$, $\mu=0.50$ and $k_s=1.0$.   
In these case, a cusp-like structure develops  around $k\approx 0.65$. 
The Euclidean-Minkowski formulation shed some light on the origin of  this cusp,
which comes actually from the last  term $\tilde{g}_i$ in equation (\ref{eq18b}).
According to its definition  in Eq. (\ref{eq19}), 
this term  does not contribute below the first meson creation threshold,
that is if \mbox{$\sqrt{s}=2\varepsilon_{k_s}<2m+\mu$}, since in this case the argument of the $\theta$-function is always negative. 

Above the threshold,  the amplitude $\tilde{g}(k)$  is also zero for $k>k_c$  where $k_c$ is a critical value given by
\begin{equation}\label{kc}
k_c=\frac{1}{2\sqrt{s}}\sqrt{(s-\mu^2)(s-(2m+\mu)^2)}
\end{equation}
When $k<k_c$, the argument of the $\theta$-function vs. the integration variable $k'$ can be positive that allows a non-zero value of $\tilde{g}_i$.
If $k\to k_c$ from below, the integration domain over $k'$  due the $\theta$-function  shrinks to 
\[ k'_c-c\sqrt{k_c-k} \leq k' \leq k'_c+c\sqrt{k_c-k}\]
where $k'_c$ and $c$ are independent of $k$ and $k'$.
One then has, in this limit, 
\[\tilde{g}_i(k)\propto \sqrt{k_c-k}\]
This is the reason of the cusp behavior, with an infinite derivative  at $k=k_c$, which manifests itself in the lower panel of Fig. \ref{Foff_ks_1.0}.  
For $m=1,\mu=0.5$ and $k_s=1$ the critical value, given by Eq. (\ref{kc}), is $k_c=0.651$, in agreement  with the position of the cusp seen in our results. 
We have also checked from our  numerical solutions,  that the cusp position 
moves as function of $k_s$ according to Eq. (\ref{kc}). 
It is interesting to point out that from the analysis of our purely Minkowski BS equation (\ref{Eq_F_sym})
the origin of this cusp  is not evident at all.
It is however remarkably manifested in the corresponding numerical solution seen in Fig.  \ref{Foff_ks_1.0}.  
We would like to emphasize that this cusp-like structure described analytically above, is only one example
of the many structures seen in the results displayed in Figs. \ref{FIG_f_k0} and \ref{FIG_f_k}, though their analysis
was beyond the scope of the present work.   

The small differences near the cusp between the real parts of  ${F_0}(k)$ and   $\tilde{F_0}(k)$  are purely numerical and indicate the difficulty in reproducing sharp behaviors in terms of smooth functions.  They can be reduced by increasing the number of  grid points.

The results presented in Fig. \ref{Foff_ks_1.0}  confirm the validity and accuracy of our direct Minkowski space calculations.

\section{Conclusion}\label{concl}

We present a new method to solve the Bethe-Salpeter equation in Minkowski space.
Contrary to the preceding approaches devoted to this problem, this method
does not make use of the Nakanishi integral representation
of the amplitude but it is based on a direct solution of the equation
taking properly into account the many singularities.
A regular equation is finally obtained and solved numerically by standard methods. 

It has been successfully applied to bound and scattering states.
The  Bethe-Salpeter off-shell scattering amplitude in Minkowski space  has been computed for the first time.

Applying the  Wick rotation to the original Bethe-Salpeter equation, an independent system of equations coupling the Euclidean amplitude to the Minkowski one for a
particular off-shell value has been derived.
It provides an independent test for our approach.

Coming on mass shell,  the elastic phase shifts and low energy parameters where accurately computed.
They considerably differ, even at zero energy, from the non-relativistic ones.
Above the meson creation threshold, an imaginary part of the phase shift  appears  and has also been  calculated.

The results presented here were limited  to S-wave in the spinless case and the ladder kernel
but they can be extended to any partial wave.

The  off-shell Bethe-Salpeter  scattering amplitude thus obtained has been further used to calculate the transition form factor \cite{transit}. 
In its full off-shell form, it can be used as  input in the three-body Bethe-Salpeter Faddeev equations.

\begin{acknowledgments}
Part of this work was done in the framework of
the  CNRS and RAS agreement involving the Lebedev Physical Institute in Moscow and the Institut
de Physique Nucleaire d'Orsay (IPNO).
One of the authors (J.C.) is grateful to the Lebedev Physical Institute, Moscow, for
being so welcome in the several visits to this laboratory.
Other author (V.A.K.) is grateful to IPN Orsay and SPhN Saclay for the kind
hospitality during his visits.

\end{acknowledgments}


\appendix
\bigskip
\section{Numerical methods}\label{Numerics}

We consider a generic two-dimensional integral equation 
\begin{widetext}
\begin{eqnarray}
f(x,y) &=&  f^B(x,y) \cr
         &+& \int_0^{\infty}  dy' V(x,y,a_{y'},y') f(a_{y'},y')   + \int_0^{\infty}  dy' V(x,y,b_{y'},y') f(b_{y'},y') \cr
         &+&   \int_0^{\infty}  dy'  \int_0^{\infty} dx' \; \frac{ V(x,y,x',y') f(x',y') - V(x,y,a_{y'},y') f( a_{y'} ,y')}{{x'}^2-a^2_{y'} }  \cr
         &+&  \int_0^{\infty}  dy'   \int_0^{\infty} dx' \; \frac{ V(x,y,x',y') f(x',y') - V(x,y,b_{y'},y') f( b_{y'} ,y')}{{x'}^2-b^2_{y'} }   \label{NUM1}
\end{eqnarray}
\end{widetext}

The solution $f$ is searched in  the compact domain $[0,x_m]\times [0,y_m]$ in the form
\begin{equation}
f(x,y) =  \sum_{ij} c_{ij} S_{i}(x)S_{j}(y)  \label{NUM2}
\end{equation}
where $c_{ij} $ are  unknown coefficients  to be determined and $S_i$ is a basis of  spline functions (see for instance \cite{ck-trento}). They are defined respectively in $[0,x_m]$ and $ [0,y_m]$
and are cubic piecewise in each of the $N_x$ ($N_y$) intervals in which $[0,x_m]$ ($ [0,y_m]$) is divided.   

The expansion (\ref{NUM2}) is supposed to be valid on a set of selected point $\{ \bar{x}_i\} \times \{ \bar{y}_j\}$ with $i=0,2N_x+1$ and   $j=0,2N_y+1$ 
suitably chosen in order to maximize the accuracy of the solution.

By inserting (\ref{NUM2}) in (\ref{NUM1}) one is led with a linear system of equations 
\begin{equation}
\sum_{i'j'} U_{ij,i'j'}  \; c_{i'j'}   = f^B_{ij} +   \sum_{i'j'} A_{ij,i'j'}   \; c_{i'j'}
\end{equation}
with an inhomogeneus term given by
\begin{equation}
f^B_{ij}   = f^B(\bar x_i,\bar y_j)   
\end{equation}
and the matrices $U$ and $A$  are
\begin{equation}
U_{ij,i'j'} = S_{i'}(\bar x_i)S_{j'}(\bar y_j)   
\end{equation}
and
\begin{widetext}
\begin{eqnarray}
A_{ij,i'j'} &=&   \int_0^{y_{max}}  dy'   \;        V(\bar x_i,\bar y_j,a,y')S_{i'}( a_{y'} )  S_{j'}(y') \cr
               &+&   \int_0^{y_{max}}  dy'   \;       V(\bar x_i,\bar y_j,b,y')S_{i'}(b)  S_{j'}(y') \cr
               &+&   \int_0^{y_{max}}  dy'   \;       S_{j'}(y') \int_0^{x_{max}} dx' \frac{V(\bar x_i,\bar y_j,x',y')S_{i'}(x') -V(\bar x_i,\bar y_j, a_{y'} ,y')S_{i'}( a_{y'} ) }{{x'}^2- a_{y'} ^2}  \cr
               &+&   \int_0^{y_{max}}  dy'   \;       S_{j'}(y') \int_0^{x_{max}} dx' \frac{V(\bar x_i,\bar y_j,x',y')S_{i'}(x') -V(\bar x_i,\bar y_j, b_{y'} ,y')S_{i'}( b_{y'} ) }{{x'}^2- b_{y'} ^2}  \label{A_ij}
\end{eqnarray}
\end{widetext}
An interesting  property of the splines used  is the fact that the functions $S_{2i}(x)$ and $S_{2i+1}(x)$  have a support limited to the two consecutive intervals $[x_{i-1},x_{i}] \cup [x_{i},x_{i+1}]$
and vanish elsewhere. 
This reduces considerably the computation of the matrix elements.

After removing the many singularities following the techniques explained in Sec. \ref{transform}, all integrands appearing in (\ref{A_ij}) are regular functions and the integrations can be performed using the standard 
Gauss quadrature methods.
The number collocation points on each dimension equals the number of spline basis and the validation procedure allows to determine the coefficients $c_{ij}$ of the expansion (\ref{NUM2}).
One is finally led to solve a complex linear system denoted symbolically
\[ (U-A ) c= f^b \]
with dimension  $d=(2N_x+1)(2N_y+1)$

\bigskip
When dealing with a finite integration domain in both variables $k',k'_0$ some care must be taken to use the subtraction technique (\ref{subtr}).
Indeed this relation -- used for $k'$ as well as for $k'_0$ integrations --  it is based on the identity (\ref{subtr1}) which is valid only in an infinite domain and that must be properly adapted. 
Thus,  for a generic variable $z=x,y$ integrated over a finite domain $z\in[0,L]$, the relation
\[  \int_0^{\infty} \frac{dz}{{z}^2-a^2}=0\]
must be replaced by
\begin{equation}\label{subtr3}
\int_0^{L} \frac{dz}{{z}^2-a^2}+\frac{1}{2
 a }\log \left|\frac{L+a}{L-a}\right|=0
\end{equation}
The integral term in (\ref{subtr3}) is used to eliminates the singularities on the finite interval $[0,L]$ 
whereas the logarithmic term represents a finite volume correction. 
 
\section{Deriving equation (\protect{\ref{subtr2}})}\label{appen1}

Let us consider the integral:
\begin{equation}\label{intx1}
I(y)= PV\int_{-\infty}^{\infty}\frac{dx}{x^2-y^2}
\end{equation}
appearing in (\ref{subtr2}).
Since it is zero if $y\neq 0$ (see Eq. (\ref{subtr1})) and diverges if $y=0$, we expect it to be proportional to the delta-function $\delta(y)$.
We replace it by the regularized integral 
\begin{equation}\label{intx2}
I_{\epsilon}(y)=\int_{-\infty}^{\infty}\frac{(x^2-y^2)dx}{(x^2-y^2)^2+\epsilon^2}
\end{equation}
which tends to $I(y)$ when $\epsilon \to 0$. Calculating this integral, we find:
$$
I_{\epsilon}(y)=\frac{\pi\sqrt{\sqrt{y^4+\epsilon^4}-y^2}}{\sqrt{2}\sqrt{y^4+\epsilon^4}}
$$
When $\epsilon \to 0$, this is a very sharp function in the vicinity of $y=0$ with the value $I_{\epsilon}(0)=\pi/(\sqrt{2}\epsilon)\to\infty$.
It represents a delta-function. The integral 
$$
\int_{-\infty}^{\infty}I_{\epsilon}(y)dy=\frac{\pi^2}{2}
$$
gives the normalization coefficient.
We conclude that
\begin{equation}\label{intx3}
I(y)= PV\int_{-\infty}^{\infty}\frac{dx}{x^2-y^2} = \frac{\pi^2}{2} \delta(y)
\end{equation}
\bigskip

\section{Coupled Euclidean-Minkowski system of equations for the \mbox{S-wave} amplitudes}\label{first}

In Sec. \ref{euclid_scat}, making a Wick rotation in the BS equation, we derived the system of equations coupling the Euclidean amplitude 
$F^E(k_4,k,z)$ and the Minkowski one for the particular off-shel $k_0$-value $\tilde{F}^M(k,z)=F^M(k_0=\varepsilon_{k_s}- \varepsilon_{k},k,z)$ with
 $k\in[0, k_s]$. To underline the main steps, this was done without partial waves decomposition.
Here we give the equations for the S-wave, which we solved numerically. 
We remind that the partial waves $F_L^E$ and $F_L^M$  are defined by eq . (\ref{fpw2}).

The first equation reads:
\begin{widetext}
\begin{eqnarray}
{F}^E_0(k_4,k)&=& \frac{1}{16\pi}V_0^{B}(k_4,k)   \nonumber\\
&+&\frac{1}{4\pi^3}\int_0^{\infty} {k'}^2 dk'\; \int_{0}^{\infty}d{k'}_4
\frac{V_s(k_4,k;k'_4,k')}{({k'_4}^2 +a_-^2)}
\left\{\frac{{F}^E_0(k',{k'}_4)}{({k'_4}^2+a_+^2)}
-
\frac{{F}^E_0({k'}_4=0,k')}{a_+^2}\right\}
\nonumber\\
&+&\frac{1}{4\pi^3}\int_0^{\infty}{k'}^2 dk'\;\pi\left\{ \frac{{F}^E_0({k'}_4=0,k')} {|a_-|a_+^2}V_1(k_4,k,k')  \right. 
  -\left.\frac{k_s}{k'}\frac{\varepsilon_{k_s}^3{F}^E_0({k'}_4=0,k_s)}{2\varepsilon_{k'}^4 |a_-|a_+}    V_0^{B}(k_4,k)\right\}   \nonumber\\
&+&   h + g   \label{eq18}
\end{eqnarray}
where the kernels 
\begin{eqnarray}
V_0^{B}(k_4,k) &=& \frac{4\pi m^2\alpha}{kk_s}\log\frac{k_4^2+\mu^2+(k+k_s)^2} {k_4^2+\mu^2+(k-k_s)^2}   \label{V00}  \\
V_s(k_4,k;k'_4,k') &=& \frac{4\pi m^2\alpha}{kk'}\log\frac{\left[k_4^2+{k'_4}^2+\mu^2+(k+k')^2\right]^2-4k_4^2{k'}_4^2}
{\left[k_4^2+{k'_4}^2+\mu^2+(k-k')^2\right]^2-4k_4^2{k'}_4^2}  \label{Vs}  \\
V_1(k_4,k,k' )&=&\frac{4\pi m^2\alpha}{ k k'}    \log\frac{\left[|a_-|+ \sqrt{(k+k')^2+\mu^2}\right]^2 +k_4^2}{\left[|a_-|+ \sqrt{(k-k')^2+\mu^2}\right]^2 +k_4^2}  \label{V1}
\end{eqnarray}

The amplitude $\tilde{F}^M_0(k)$ enters in the terms $h$ and  $h$.

The term $h$ is given by:
\begin{eqnarray}\label{hs}
h(k_4,k)&=&\frac{1}{16\pi^2} F^E_0({k'}_4=0,k=k_s)V_0^{B}(k_4,k)
\frac{k_s}{\varepsilon_{k_s}}\left(\frac{\varepsilon_{k_s}^2}{m^2} +4\log\frac{\varepsilon_{k_s}}{m}-2-\log 4\right)
\nonumber\\
&+&\frac{1}{8\pi^2}\tilde{F}^M_0(k=k_s)V_0^{B}(k_4,k)
\frac{1}{\varepsilon_{k_s}}\left(2m\arctan\frac{k_s}{m}-k_s\log\frac{2k_s}{m} \right)
\end{eqnarray}
\end{widetext}
It does not contain the integrals and it contains the on-shell Minkowski and Euclidean amplitudes which coincide with each other.

The term $g$ is given by:
\begin{widetext}
\begin{eqnarray}
g(k_4,k)&=&\frac{1}{8\pi^2} \int_0^{k_s} {k'}^2 dk'
\left\{\frac{\tilde{F}^M_0(k')}{ \varepsilon_{k'}\varepsilon_{k_s}a_-}\right.
V_2(k_4,k,k')
-\left.\frac{k_s}{k'}\frac{2\varepsilon_{k_s}\tilde{F}^M_0(k=k_s)}
{\varepsilon_{k'}^2a_+a_-}
V_0^{B}(k_4,k)\right\}
\label{Res}
 \cr
&+&
\frac{i}{16\pi}\frac{k_s}{\varepsilon_{k_s}} \tilde{F}^M_0(k=k_s)\;V_0^{B}(k_4,k)
\end{eqnarray}
It contains  on the kernel $V_2(k_4,k,k')$:
\begin{equation}\label{V2}
V_2(k_4,k,k')=\frac{2\pi m^2\alpha}{kk'}\log\left|\frac{[k_4^2+a_-^2+ \mu^2+(k+k')^2]^2-4a_-^2[\mu^2+(k+k')^2]}
{[k_4^2+a_-^2+ \mu^2+(k-k')^2]^2-4a_-^2[\mu^2+(k-k')^2]}\right|
\end{equation}
\end{widetext}
Notice that:
$$
V_1(k_4,k,k'=k_s)=V_2(k_4,k,k'=k_s)=V_0^{B}(k_4,k)
$$
where $V_0^{B}(k,k_4)$ is defined in (\ref{V00}).

This completes the full definition of the equation (\ref{eq18}). Due to subtraction which we have under any integral, these integrals are non-singular.

The second equation has the following form:
\begin{widetext}
\begin{eqnarray}
&&\tilde{F}^M_0(k)= \frac{1}{16\pi}V_0^{B}\Bigl(k_4=i(\varepsilon_{k_s}-\varepsilon_{k}),k\Bigr)   \cr
&&+\frac{1}{4\pi^3}\int_0^{\infty} {k'}^2 dk'\; \int_{0}^{\infty}d{k'}_4
\frac{V_s\Bigl(k_4=i(\varepsilon_{k_s}-\varepsilon_{k}),k;k',k'_4\Bigr)}
{({k'_4}^2 +a_-^2)}
\left\{\frac{{F}^E_0({k'}_4,k')}{({k'_4}^2+a_+^2)}
-
\frac{{F}^E_0({k'}_4=0,k')}{a_+^2}\right\}  Ê\cr
&+&\frac{\pi}{4\pi^3}\int_0^{\infty}{k'}^2 dk'\left\{
\frac{{F}^E_0({k'}_4=0,k')} {|a_-|a_+^2}
V_1\Bigl(k_4=i(\varepsilon_{k_s}-\varepsilon_{k}),k,k'\Bigr)
\frac{k_s\varepsilon_{k_s}^3{F}^E_0({k'}_4=0,k_s)     }{2k'\varepsilon_{k'}^4 |a_-|a_+}
V_0^{B}\Bigl(k_4=i(\varepsilon_{k_s}-\varepsilon_{k}),k\Bigr)     \right\}  \cr
&+&   \tilde{h} +  \tilde{g}  +\tilde{g}_i   \label{eq18b}
\end{eqnarray}
\end{widetext}

In contrast to Eq. (\ref{eq18}), the equation (\ref{eq18b}) contains the term 
\begin{widetext}
\begin{equation}\label{eq19}
\tilde{g}_i (k) = \frac{ig^2}{64\pi}\int_0^{k_s}\frac{{k'}^2dk'}{kk'\varepsilon_{k_s} \varepsilon_{k'}a_-}
\tilde{F}^M_0(k')\;
\theta\left(1-\frac{|(2\varepsilon_{k_s}-\varepsilon_{k'} -\varepsilon_{k})^2-{k'}^2-k^2-\mu^2|}{2kk'}\right)
\end{equation}
\end{widetext}
Due to restriction given by the theta-function, one can show that this term is identically zero below the  one meson creation  threshold $2m+\mu$. 
Namely it provides the value $Im[\delta]$ above the threshold: if we omit it we  find always $Im[\delta]=0$. 
The denominator \mbox{$a_-=\varepsilon_{k'}-\varepsilon_{k_s}$} never crosses zero since the theta-function does not allow that (it restricts the domain where $a_-\neq 0$).
Kernels $V_0^{B}$, $V_s$, $V_1$  and $V_2$ are correspondingly defined  by eqs. (\ref{V00}), (\ref{Vs}), (\ref{V1}) and (\ref{V2})  above. 

The quantity $\tilde{h}$  reads:
\begin{widetext}
\begin{eqnarray}\label{hsb}
&&\tilde{h}(k)
= \frac{1}{16\pi^2}{F}^E_0({k}_4=0,k=k_s)
V_0^{B}\Bigl(k_4=i(\varepsilon_{k_s}-\varepsilon_{k}),k\Bigr)
\frac{k_s}{\varepsilon_{k_s}}\left(\frac{\varepsilon_{k_s}^2}{m^2} +4\log\frac{\varepsilon_{k_s}}{m}-2-\log 4\right)
\nonumber\\
&&+\frac{1}{8\pi^2}\tilde{F}^M_0(k=k_s)
V_0^{B}\Bigl(k_4=i(\varepsilon_{k_s}-\varepsilon_{k}),k\Bigr)
\frac{1}{\varepsilon_{k_s}}\left(
2m\arctan\frac{k_s}{m}-k_s\log\frac{2k_s}{m} \right)
\end{eqnarray}

The term $\tilde{g}$  has the form:
\begin{eqnarray}
\tilde{g}(k) &=&\frac{1}{8\pi^2} \int_0^{k_s} {k'}^2 dk'
\left\{\frac{\tilde{F}^M_0(k')}{ \varepsilon_{k'}\varepsilon_{k_s}a_-}\right.
V_2\Bigl(k_4=i(\varepsilon_{k_s}-\varepsilon_{k}),k,k'\Bigr)
\nonumber
-\left.\frac{k_s}{k'}\frac{2\varepsilon_{k_s}\tilde{F}^M_0(k=k_s)}
{\varepsilon_{k'}^2a_-a_+}
V_0^{B}\Bigl(k_4=i(\varepsilon_{k_s}-\varepsilon_{k}),k\Bigr)\right\}
\label{Resb}
\\
&+&\frac{i}{16\pi}\frac{k_s}{\varepsilon_{k_s}} \tilde{F}^M_0(k=k_s)\;
V_0^{B}\Bigl(k_4=i(\varepsilon_{k_s}-\varepsilon_{k}),k\Bigr)
\label{Imsb}
\end{eqnarray}
\end{widetext}


\end{document}